\documentclass{article} 
\usepackage{iclr2027_conference,times}

\usepackage{amsmath,amsfonts,bm}

\def\eqref#1{equation~\ref{#1}}

\def\1{\bm{1}}

\DeclareMathAlphabet{\mathsfit}{\encodingdefault}{\sfdefault}{m}{sl}
\SetMathAlphabet{\mathsfit}{bold}{\encodingdefault}{\sfdefault}{bx}{n}

\usepackage{hyperref}
\usepackage{url}
\usepackage{graphicx}
\usepackage{booktabs}
\usepackage{pifont}
\usepackage{multirow}
\usepackage{amssymb}

\hypersetup{
    colorlinks=true,
    citecolor=blue,
    linkcolor=black,
    urlcolor=blue
}

\title{\raggedright MeshSuite: A Unified Benchmark for Triangle- and Mesh-Based Neural Rendering}

\iclrfinalcopy
\author{%
  \textbf{Kaixuan Zhang}$^{1,2}$
  \quad
  \textbf{Minxian Li}$^1$\thanks{Corresponding author.}
  \quad
  \textbf{Mingwu Ren}$^1$
  \quad
  \textbf{Xiatian Zhu}$^2$
  \vspace{.5em} 
  \\
  $^1$Nanjing University of Science and Technology
  \qquad
  $^2$University of Surrey
}

\begin{document}

\maketitle

\begin{abstract}
Triangle- and mesh-based neural rendering aims to bridge neural scene representations and existing graphics engines (\textit{e.g.}, Unity, Blender) by leveraging \textit{triangle} primitives compatible with standard rasterization hardware.
There have been several such methods driven by parallel efforts, developed and evaluated under inconsistent settings, with little or no comparison with each other; critically, most even have never used graphics engines for evaluation nor considered deployability in practice -- significantly undermining the objective and motivation.
To address these issues, we introduce \textbf{MeshSuite}, the first benchmark of its kind to enable systematic evaluation of graphics-engine deployment and the \texttt{de facto renderer} for triangle- and mesh-based neural rendering methods.
Importantly, our deployment protocol includes two engine deployment settings: 
(1) \texttt{standard engine deployment} -- a conventional opaque mesh pipeline with vertex colors and hardware Z-buffering;
(2) \texttt{custom engine deployment} -- adding method-specific engine implementations supporting the retained appearance and compositing features (\textit{e.g.}, alpha blending), so that the characteristics of each specific model can be taken into account.
For mesh splatting, we further propose a structural audit of exported surfaces, diagnosing topological and geometric integrity toward downstream graphics assets. 
We highlight several results:
(1) Standard engine deployment incurs image-quality degradation for every method and training configuration studied. Opaque designs degrade the least, including configurations trained for engine use;
(2) The deployment setting matters -- custom engine deployment matches the de facto renderers for all methods on Mip-NeRF~360, at a roughly 3$\sim$7$\times$ reduction in throughput relative to standard engine deployment for the alpha-blended methods;
(3) About mesh splatting, the current approaches to explicit connectivity and shared indexing are still limited: shared vertex indexing alone does not ensure manifoldness or global connectivity.
From this benchmark, we validate that \textit{rasterizability is merely part of graphics readiness}, and highlight the significance of assessing the graphics engine deployment process (\textit{e.g.}, the degree of engine compatibility). 
Project page: \url{https://x-up-lab.github.io/research/meshsuite}.
\end{abstract}

\section{Introduction}
\label{sec:intro}
Reconstructing a scene from multi-view images allows users to explore it from arbitrary viewpoints beyond those originally captured, supporting a wide range of vision and graphics applications, including AR/VR \citep{broxton2020immersive}, gaming \citep{unity_photogrammetry}, and robotic perception \citep{nice_slam}. Recent advances in neural rendering, particularly Neural Radiance Fields (NeRFs) \citep{nerf} and 3D Gaussian Splatting (3DGS) \citep{3dgs}, have enabled increasingly realistic novel-view synthesis (NVS). Specifically, NeRFs use multilayer perceptrons (MLPs) to encode scene density and view-dependent color, synthesizing novel views through volume rendering \citep{nerf}. However, their practical use is constrained by high computational costs and the lack of an explicit geometric representation. In contrast, 3DGS represents scenes explicitly with 3D Gaussian primitives and uses efficient differentiable splatting to substantially reduce rendering costs \citep{3dgs}. 

Despite these advances, neither representation is directly compatible with the standard rendering pipelines of established graphics engines, such as Blender \citep{blender} and Unity \citep{unity}. These engines have benefited from decades of development and optimization, providing efficient rendering pipelines and integrated tools for lighting and shadows, material editing, and animation. Their mature ecosystems also offer extensive collections of reusable assets and plugins \citep{unity, blender}, facilitating the integration of reconstructed scenes into practical applications.

To better take advantage of these engines, researchers have increasingly explored triangle- and mesh-based neural rendering methods, which optimize triangle primitives already supported by standard graphics hardware \citep{trianglesplatting, meshsplatting}. These methods have demonstrated impressive NVS performance comparable to that of NeRFs and 3DGS, while offering a promising path towards integration with existing graphics engines \citep{trianglesplatting, 2dts, meshsplatting}. 
However, existing triangle- and mesh-based works follow separate evaluation protocols, with differences in data preprocessing, image formats, evaluation metrics, and metric implementations. Importantly, none of these works evaluates their 3D models in a graphics engine, still using their \texttt{de facto renderers}, typically realized by CUDA-based differentiable rasterizers \citep{meshsplatting, diffsoup, trianglesplatting, 2dts}.

Using triangles alone also does not ensure complete engine compatibility. Existing methods use triangles with substantially different features (see Tab. \ref{tab:design_features}), including continuous or discrete opacity \citep{trianglesplatting, meshsplatting}, coverage defined inside or outside a triangle \citep{2dts, trianglesplatting}, varying color representations (\textit{e.g.}, view-dependent spherical harmonics (SHs) \citep{meshsplatting} or neural networks \citep{diffsoup}), and the presence or absence of neural texture decoders \citep{diffsoup, trianglesplatting}. These variations pose challenges for engine rendering and may require method-specific shaders, buffers, sorting strategies, and runtime state. 
Furthermore, existing mesh splatting methods \citep{meshsplatting} evaluate exported mesh quality solely through vertex sharing, lacking a comprehensive analysis of critical structural and topological properties such as manifoldness and global connectivity.

\begin{figure}[t]
\centering
\includegraphics[width=\textwidth]{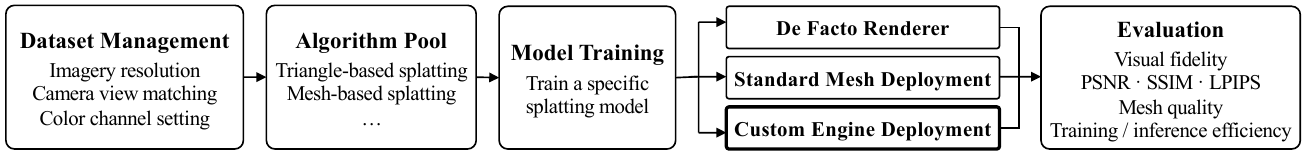}
\caption{\textbf{MeshSuite:} The \texttt{Dataset Management} layer standardizes data formats, including image resolution, camera-view alignment, and color-channel settings. Triangle- and mesh-based splatting methods are integrated into the \texttt{Algorithm Pool} layer and trained through the \texttt{Model Training} layer.
The trained model can be evaluated in three ways: de facto renderer, standard engine deployment, and custom engine deployment; the latter two use a chosen graphics engine. Its performance is then evaluated in the \texttt{Evaluation} layer using a wide range of metrics covering visual fidelity, mesh structure, and training/inference efficiency.
}
\label{fig:benchmark_protocol_teaser}
\end{figure}

To address these questions, we introduce \textbf{MeshSuite}, a unified benchmark for evaluating triangle- and mesh-based neural rendering (see Fig.~\ref{fig:benchmark_protocol_teaser}).
This benchmark enables: 
(a) integration of existing triangle- and mesh-based splatting methods; 
(b) fair and consistent comparison of different methods in a unified setting; 
(c) extensive evaluation of both triangle- and mesh-based models during training and inference; 
(d) de facto renderer evaluation and two engine deployment settings:
\texttt{standard engine deployment} -- converting a trained model to a standard opaque vertex-colored mesh before rendering in a graphics engine;  
\texttt{custom engine deployment} -- adding a customized layer to implement method-specific designs, such as specialized appearance representations and alpha blending;
(e) systematic quality evaluation of meshes exported by mesh splatting methods.

We note three key findings. \textit{First}, standard engine deployment reduces visual quality relative to the de facto renderer for every method and training configuration, with opaque designs losing the least: DiffSoup \citep{diffsoup} with baked appearance, MeshSplatting \citep{meshsplatting}, and 2DTS trained for opaque rendering. The alpha-blended triangle soups trained for novel-view synthesis lose more than 12~dB PSNR (Tab.~\ref{tab:mip360_renderer_comparison}).
\textit{Second}, custom engine deployment matches the de facto renderer to within 0.05~dB PSNR for all methods on Mip-NeRF~360, at the cost of a roughly 3$\sim$7$\times$ reduction in throughput for the alpha-blended methods (Tab.~\ref{tab:mip360_renderer_comparison}).
\textit{Third}, existing mesh splatting methods' exported meshes exhibit non-manifold structures and substantial fragmentation (Tab.~\ref{tab:mesh_connectivity}).
These results show that \textit{using triangles alone does not ensure graphics readiness}; appearance preservation, efficiency, and mesh structure must be evaluated together.

Our \textbf{\textit{contributions}} are as follows: (\textbf{I}) the first unified benchmark, \textit{MeshSuite}, for evaluating triangle- and mesh-based neural rendering methods; (\textbf{II}) custom engine deployment of existing methods in a graphics engine (\textit{i.e.}, Unity), retaining their specific designs during engine rendering; and (\textbf{III}) an extensive evaluation of existing models in terms of novel-view synthesis, training/inference cost, and mesh structure.

\section{Related Work}

\begin{table}[t]
\centering
\footnotesize
\resizebox{\linewidth}{!}{%
\begin{tabular}{@{}l|cc|ccc|ccc@{}}
\toprule
\multirow{2.5}{*}{\textbf{Method}} & \multicolumn{2}{c|}{\textbf{Geometry \& Topology}} & \multicolumn{3}{c|}{\textbf{Coverage \& Opacity}} & \multicolumn{3}{c}{\textbf{Rendering \& Appearance}} \\
\cmidrule(lr){2-3}\cmidrule(lr){4-6}\cmidrule(l){7-9}
 & \textbf{Primitive} & \textbf{Shared Vert.} & \textbf{Bounded} & \textbf{Soft Cov.} & \textbf{Opacity} & \textbf{Depth Sort} & \textbf{Appearance} & \textbf{Neural Dec.} \\
\midrule
2DTS & Triangle soup & \ding{55} & \ding{55} & \ding{51} & Continuous & \ding{51} & View-dep.\ SH & \ding{55} \\
Triangle-Splatting & Triangle soup & \ding{55} & \ding{51} & \ding{51} & Continuous & \ding{51} & View-dep.\ SH & \ding{55} \\
MeshSplatting & Indexed mesh & \ding{51} & \ding{51} & \ding{55} & Binary & \ding{51} & View-dep.\ SH & \ding{55} \\
DiffSoup & Triangle soup & \ding{55} & \ding{51} & \ding{51} & Binary & \ding{55} & Neural texture & \ding{51} \\
\bottomrule
\end{tabular}}
\caption{Key design features of evaluated methods. \emph{Shared Vert.}: indexed mesh with shared vertices and explicit topological connectivity; \emph{Bounded}: primitive coverage strictly confined within the geometric face boundary; \emph{Soft Cov.}: smooth spatial opacity falloff toward primitive boundaries; \emph{Depth Sort}: requires primitive-level depth sorting for alpha compositing. \emph{Neural Dec.}: requires a neural decoder at render time.}
\label{tab:design_features}
\end{table}

\textbf{Dominant neural representations.} Novel-view synthesis (NVS) aims to synthesize images of a scene from viewpoints beyond those observed in the input images. A central component of this task is a scene representation that captures both geometry and appearance. NeRF represents a scene as a continuous function and synthesize views through volumetric integration \citep{nerf}. Later methods improve reconstruction quality and anti-aliasing \citep{mipnerf,mipnerf360,zipnerf} or accelerate rendering through explicit, factorized, and hashed structures \citep{kilonerf, plenoctrees, plenoxels, tensorf, instantngp}. 3DGS replaces repeated network queries with optimized anisotropic Gaussian primitives and a dedicated splatting rasterizer \citep{3dgs}. Follow-up works refine Gaussian optimization or encourages surface-aligned geometry \citep{3dgs-mcmc, 2dgs, sugar}. These methods establish strong NVS baselines, but their representations are not natively supported by conventional graphics engines (\textit{e.g.}, Unity and Blender). Integration with existing graphics engines therefore requires specialized rendering implementations or conversion to mesh-based assets.

To facilitate integration with conventional graphics engines, methods such as MobileNeRF \citep{mobilenerf} and BakedSDF \citep{bakedsdf} convert learned radiance fields into mesh-based representations.
For explicit splats, SuGaR \citep{sugar} and GaMeS \citep{games} extract mesh surfaces from 3D Gaussian primitives or bind Gaussian primitives to deformable mesh faces. Although these conversion pipelines produce standard mesh assets, they often rely on lossy post-processing (\textit{e.g.}, Marching Cubes, Poisson surface reconstruction, or heuristic simplification) that decouples the optimization target from the final rendering objective. 

\textbf{Triangle- and mesh-based neural representations.}
Existing graphics engines (\textit{e.g.}, Blender \citep{blender} and Unity \citep{unity}) provide optimized rendering pipelines and integrated tools for scene editing, lighting, and animation, together with mature ecosystems of reusable assets and plugins. Triangle meshes are widely supported by these pipelines, making them a natural choice for bringing learned scene representations into existing graphics workflows.

To exploit this support, triangle- and mesh-based neural rendering methods optimize geometry directly represented by triangle primitives \citep{trianglesplatting,2dts,meshsplatting,diffsoup}. These approaches use either \emph{triangle soups} with independent faces or \emph{indexed meshes} with shared vertices and explicit connectivity. 2DTS \citep{2dts} and Triangle-Splatting \citep{trianglesplatting} optimize independent triangles with soft coverage and alpha blending, DiffSoup \citep{diffsoup} uses neural texture features and a color decoder, MeshSplatting \citep{meshsplatting} introduces shared-vertex connectivity. Tab. \ref{tab:design_features} summarizes and contrasts the key design features across these representations. These works demonstrate the potential of triangle- and mesh-based representations for high-quality NVS and offer promising routes toward graphics-engine integration.

However, using triangle primitives alone does \textit{not} ensure complete engine compatibility, as method-specific appearance and transparency models may still require additional adaptation. These methods are also developed and reported under different evaluation pipelines, making direct cross-method comparison difficult. Their original evaluations primarily use their de facto renderers rather than assess the learned representations after deployment in a graphics engine. Instead, MeshSuite addresses both gaps through unified evaluation and explicit graphics-engine deployment.

\textbf{Neural rendering benchmarks.} Existing frameworks have sought to improve the reproducibility and comparability of NVS. Nerfstudio \citep{nerfstudio} and GauStudio \citep{gaustudio} provide modular research frameworks for NeRF and Gaussian Splatting, while gsplat \citep{gsplat} offers an efficient Gaussian rasterization library with extensive performance evaluations. More directly, NerfBaselines \citep{nerfbaselines} standardizes datasets, evaluation protocols, and metrics across NeRF and 3DGS methods while preserving their original implementations. Splatwizard \citep{splatwizard} further provides a specialized benchmark for 3DGS compression, evaluating rendering fidelity, computational efficiency, and geometric accuracy. The nvs-bench project \citep{nvsbench} facilitates reproducible NVS comparisons across different representations, including Triangle Splatting \citep{trianglesplatting}. Complementary studies investigate perceptual rendering quality \citep{liang2024perceptual} and robustness under corrupted observations \citep{wang2024robustness}.

Despite these efforts, existing benchmarks primarily focus on reconstruction quality and rendering efficiency within methods' original pipelines, leaving graphics-engine deployment largely unexplored. MeshSuite addresses this gap by extending the evaluation of triangle- and mesh-based neural rendering to graphics-engine deployment, with a structural assessment of reconstructed meshes.

\section{MeshSuite Framework}
\label{sec:benchmark_design}

MeshSuite provides an end-to-end benchmark spanning evaluation with the de facto renderers and deployment in graphics engines. As illustrated in Fig.~\ref{fig:benchmark_protocol_teaser}, it evaluates each trained representation under matched camera views, backgrounds, and evaluation metrics across three settings: de facto renderer, standard engine deployment, and custom engine deployment. Comparisons across these settings reveal how graphics-engine deployment affects visual fidelity and rendering efficiency. For exported indexed meshes, we also assess structural usability, including local/global connectivity and manifoldness. App.~\ref{app:software_architecture} details MeshSuite's software architecture.

\subsection{Dataset Management}
\label{sec:protocol_standardization}
Fair cross-method comparisons require consistent data and evaluation settings \citep{nerfbaselines,splatwizard}. The \texttt{Dataset Management} layer in Fig.~\ref{fig:benchmark_protocol_teaser} addresses three aspects:

\textbf{Imagery resolution.} MeshSuite manages image resolution and downsampling during data preparation, retaining each method's resolution requirements (Sec.~\ref{sec:experiments_setup}), which avoids inconsistent image downsampling operations during training and inference phases.

\textbf{Camera view matching.} For a held-out camera set $\mathcal{C}$, the benchmark standardizes camera coordinate frames, optical axes, field of view, and projection matrices, aligning the evaluated viewpoints across rendering settings.

\textbf{Color channel setting.} MeshSuite uses RGB images for image-quality evaluation and the alpha channel of RGBA images for background compositing on masked objects. Prior works adopt inconsistent backgrounds: for example, on DTU \citep{dtu}, Triangle-Splatting \citep{trianglesplatting} uses an unmasked black background, whereas 2DTS uses object masks and a white background. We standardize all methods to the white-background protocol for masked objects. Using the RGBA alpha channel as an opacity matte, both ground-truth target images and novel-view renderings are composited as
\begin{equation}
\mathbf{I}_{\mathrm{white}} = \mathbf{I}_{\mathrm{RGB}} \odot \boldsymbol{\alpha} + (1-\boldsymbol{\alpha})\cdot\mathbf{1}.
\label{eq:white_background}
\end{equation}
Here, $\mathbf{I}_{\mathrm{RGB}}$ and $\mathbf{I}_{\mathrm{white}}$ denote the input RGB and composited images, respectively; $\boldsymbol{\alpha}$ is the per-pixel opacity matte with values in [0, 1]; $\odot$ denotes element-wise multiplication, with $\boldsymbol{\alpha}$ applied to all RGB channels; and $\mathbf{1}$ is an all-ones image. The renderer background is also set to white, giving all methods the same evaluation background.

\subsection{Algorithm Pool and Model Training}
\label{sec:algorithm_pool_training}
The \texttt{Algorithm Pool} layer integrates triangle- and mesh-based splatting methods, including 2DTS \citep{2dts}, Triangle-Splatting \citep{trianglesplatting}, MeshSplatting \citep{meshsplatting}, and DiffSoup \citep{diffsoup}. These methods use either independent triangle soups or indexed meshes with shared vertices and explicit connectivity.

The \texttt{Model Training} layer follows each method's original training pipeline while applying the shared data and evaluation protocols. Each trained checkpoint can then be evaluated with its de facto renderer or translated into an asset for graphics-engine deployment. Sec.~\ref{sec:experiments_setup} provides the evaluated methods, datasets, and hardware settings.

\subsection{Rendering and Graphics Engine Deployment}
\label{sec:deployment_task}

\begin{table}[t]
\centering
\footnotesize
\setlength{\tabcolsep}{2.2pt}
\renewcommand{\arraystretch}{1.15}
\resizebox{\linewidth}{!}{%
\begin{tabular}{@{}l|cc|cccccc@{}}
\toprule
\multirow{2}{*}{\textbf{Method}}& \multicolumn{2}{c|}{\textbf{Standard engine deployment}} & \multicolumn{6}{c}{\textbf{Custom engine deployment}} \\
 & \textbf{Support} & \textbf{Appearance} & \textbf{Support} & \textbf{View-dep.} & \textbf{Opacity} & \textbf{Soft cov.} & \textbf{Depth sort} & \textbf{Appearance} \\
\midrule
3DGS & \multicolumn{2}{c|}{\textsc{N/A}} & \multicolumn{6}{c}{\textsc{N/A}} \\
2DGS & Fallback & baked color & \multicolumn{6}{c}{\textsc{N/A}} \\
\midrule
2DTS & Fallback & baked SH-DC & Method-aware & \ding{51} & \ding{51} & \ding{51} & \ding{51} & SH \\
Triangle-Splatting & Fallback & baked SH-DC & Method-aware & \ding{51} & \ding{51} & \ding{51} & \ding{51} & SH \\
MeshSplatting & Fallback & baked SH-DC & Method-aware & \ding{51} & \ding{51} & -- & \ding{51} & SH \\
DiffSoup & Fallback & baked atlas & Method-aware & -- & -- & -- & -- & MLP \\
\bottomrule
\end{tabular}}
\caption{Engine support and retained rendering features. \emph{Fallback}: static-color rendering without alpha blending (baked vertex colors or, for DiffSoup, a baked RGBA texture atlas with alpha cutout); \emph{Method-aware}: a dedicated renderer supporting the listed features. SH-DC denotes the constant SH color component. \ding{51}/--: feature retained/not used. \textsc{N/A}: unsupported.}
\label{tab:feature_mapping}
\end{table}

As illustrated in Fig.~\ref{fig:benchmark_protocol_teaser}, MeshSuite organizes rendering into three evaluation settings: de facto renderer, standard engine deployment, and custom engine deployment. The de facto renderer serves as the baseline for the two engine-deployment settings.

\textbf{Translation layer and asset interface.}
\label{sec:translation_layer}
Existing triangle- and mesh-based methods encode geometry, appearance, and opacity in method-specific ways. For example, 2DTS \citep{2dts} and Triangle-Splatting \citep{trianglesplatting} use independent triangles with learned opacity, MeshSplatting \citep{meshsplatting} uses shared vertices with view-dependent appearance, and DiffSoup \citep{diffsoup} represents appearance through learned features and a neural decoder. Engine deployment therefore requires method-specific interpretation of the trained model's parameters.

To support deployment through a common interface, MeshSuite introduces a \textit{translation layer} implemented by an adapter $\mathcal{E}_m$ for method $m$. For each trained model $\mathcal{S}_m$, the adapter maps its parameters to an engine asset $\mathcal{A}_m$ that follows the unified interface:
\begin{equation}
\mathcal{S}_{m} \xrightarrow{\mathcal{E}_{m}} \mathcal{A}_{m}.
\end{equation}
The asset covers geometry, appearance, and opacity for different mesh structures. Translation decodes the trained model before deployment, allowing the engine to consume the exported asset independently of the training framework. The asset supplies the geometry and appearance data used by standard engine deployment and custom engine deployment (see App.~\ref{app:software_architecture}).

\textbf{De facto renderer ($\mathbf{I}^{\mathrm{src}}_m$).} We render each trained checkpoint $\mathcal{S}_m$ with its de facto renderer $\mathcal{R}^{\mathrm{src}}_m$: $\mathbf{I}^{\mathrm{src}}_m(c) = \mathcal{R}^{\mathrm{src}}_m(\mathcal{S}_m, c), \forall c \in \mathcal{C}$, where $c$ denotes a camera in the held-out set $\mathcal{C}$ and $\mathbf{I}^{\mathrm{src}}_m(c)$ is the corresponding rendered image. This establishes baseline image quality before engine deployment.

\textbf{Standard engine deployment ($\mathbf{I}^{\mathrm{std}}_m$).} We render representations as conventional opaque meshes with static vertex colors and hardware Z-buffering. This setting omits method-specific shaders, neural decoders, and alpha blending. DiffSoup stores appearance in neural texture features rather than vertex colors, so its standard engine deployment instead uses a baked RGBA texture atlas with alpha cutout.

\textbf{Custom engine deployment ($\mathbf{I}^{\mathrm{cust}}_m$).} We implement method-specific Unity renderers under the same evaluation protocol that retain each method's appearance and compositing features (Tab.~\ref{tab:feature_mapping}). Comparing them with the de facto renderer and standard engine deployment reveals how much quality these features preserve and their runtime cost (Tab.~\ref{tab:mip360_renderer_comparison}).

\textbf{Differences from de facto renderers.} Engine renderers and de facto renderers can differ in primitive ordering, coverage, and attribute evaluation. Our custom renderers follow each source kernel in the details that affect image quality, such as the SH viewing direction, coverage window, and primitive culling of 2DTS and Triangle-Splatting, and the visibility rule and supersampling of MeshSplatting (App. Sec.~\ref{app:custom_renderers}).

\subsection{Evaluation}
\label{sec:deployment_gap}
The \texttt{Evaluation} layer in Fig.~\ref{fig:benchmark_protocol_teaser} assesses visual fidelity, mesh quality, and training/inference efficiency.

\subsubsection{Visual Fidelity}
\label{sec:nvs_quality}

\textbf{Metric standardization.} The \texttt{Evaluation} layer in Fig.~\ref{fig:benchmark_protocol_teaser} uses the same implementations of PSNR \citep{psnr}, SSIM \citep{ssim}, and LPIPS \citep{lpips} for all methods. LPIPS uses a VGG \citep{vgg} backbone with images normalized to [-1, 1].

\textbf{Visual fidelity and deployment gaps.} For an image-quality metric $q$ (\textit{e.g.}, PSNR, SSIM, or LPIPS), we compute the mean score $Q_m^r$ over the held-out camera set $\mathcal{C}$:
\begin{equation}
Q^{r}_m = \frac{1}{|\mathcal{C}|} \sum_{c\in \mathcal{C}}q(\mathbf{I}_m^r(c), \mathbf{I}^{\mathrm{gt}}(c)),
\qquad r\in\{\mathrm{src},\mathrm{std},\mathrm{cust}\},
\end{equation}
where $m$ indexes methods, $c$ denotes a held-out camera, and $|\mathcal{C}|$ is the number of evaluated cameras. The images $\mathbf{I}_m^r(c)$ and $\mathbf{I}^{\mathrm{gt}}(c)$ are the rendering of method $m$ under setting $r$ and the corresponding ground truth, respectively. The setting labels $\mathrm{src}$, $\mathrm{std}$, and $\mathrm{cust}$ denote de facto renderer, standard engine deployment, and custom engine deployment, respectively.

Then we define three pairwise gaps:
\begin{equation}
\Delta Q^{\mathrm{adapt}}_m = \left| Q^{\mathrm{src}}_m - Q^{\mathrm{cust}}_m \right |,\quad
\Delta Q^{\mathrm{port}}_m = \left| Q^{\mathrm{cust}}_m - Q^{\mathrm{std}}_m \right |,\quad
\Delta Q^{\mathrm{deploy}}_m = \left| Q^{\mathrm{src}}_m - Q^{\mathrm{std}}_m \right |.
\label{eq:deployment_gaps}
\end{equation}
The \emph{adaptation gap $\Delta Q^{\mathrm{adapt}}_m$} measures the change from the de facto renderer to custom engine deployment; the \emph{portability gap $\Delta Q^{\mathrm{port}}_m$} measures the additional change under standard engine deployment; and the \emph{deployment gap $\Delta Q^{\mathrm{deploy}}_m$} measures the overall change from the de facto renderer to standard engine deployment. The gaps report absolute differences and the metric values indicate whether quality improves or declines.

\subsubsection{Mesh Quality}

\textbf{Surface geometry.} We evaluate Chamfer Distance (CD) on DTU, which  adopts standard pipelines to isolate the object surface, complementing the white-background image protocol in Sec.~\ref{sec:protocol_standardization}.

\begin{figure}[t]
\centering
\includegraphics[width=\linewidth]{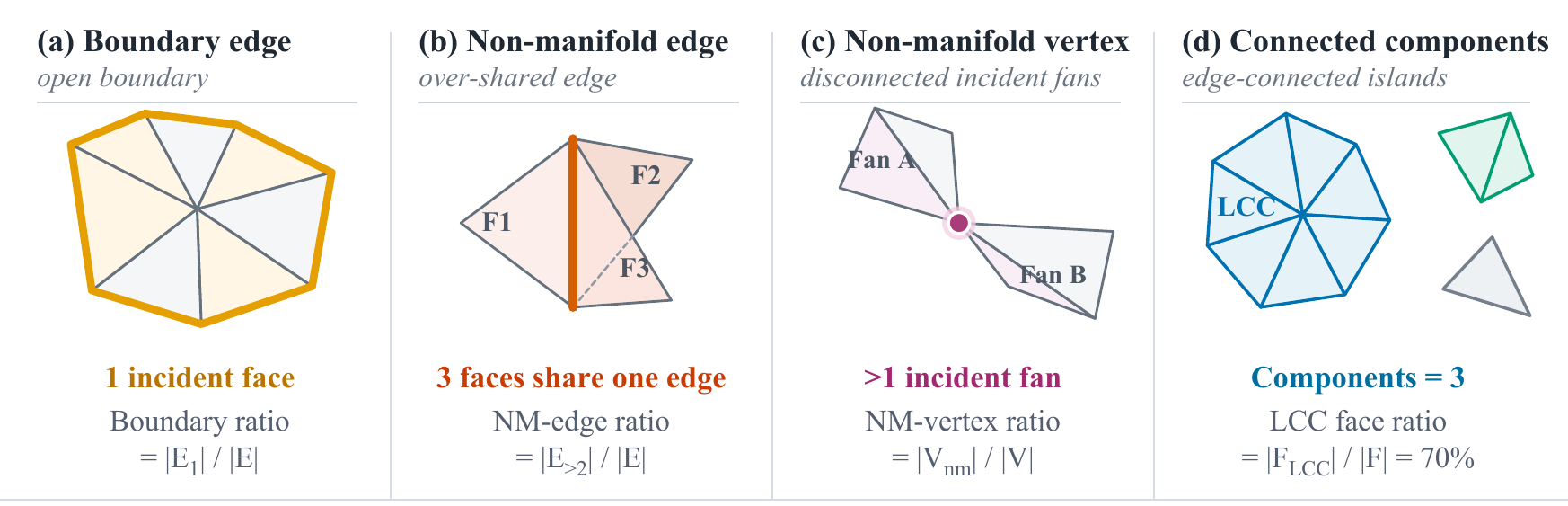}
\caption{Topology diagnostics for the structural audit. (a) Boundary edges have one incident face. (b) Non-manifold edges have more than two incident faces. (c) Non-manifold vertices have disconnected incident-face fans. (d) Edge-connected components are maximal face sets connected through shared edges.}
\label{fig:topology_diagnostic}
\end{figure}

\textbf{Mesh structural integrity.} Rendering as triangles does not guarantee a valid geometric surface. Let $\mathbf{V}$, $\mathbf{E}$, and $\mathbf{F}$ denote the sets of unique vertices, undirected edges, and triangular faces in an exported mesh, with $|\cdot|$ denoting set cardinality. We audit three complementary aspects:

\begin{itemize}
    \item \emph{Local reuse.} The ratio $|\mathbf{V}|/|\mathbf{F}|$ measures unique vertices per face, and average valence $2|\mathbf{E}|/|\mathbf{V}|$ summarizes local adjacency density. With a common export convention, lower $|\mathbf{V}|/|\mathbf{F}|$ indicates greater vertex sharing.

    \item \emph{Local structure.} The boundary-edge and non-manifold-edge ratios are $|\mathbf{E}_1|/|\mathbf{E}|$ and $|\mathbf{E}_{>2}|/|\mathbf{E}|$, where $\mathbf{E}_1$ and $\mathbf{E}_{>2}$ contain edges incident to one and more than two faces, respectively. The non-manifold-vertex ratio is $|\mathbf{V}_{\mathrm{nm}}|/|\mathbf{V}|$, where $\mathbf{V}_{\mathrm{nm}}$ is the set of vertices whose incident faces fail to form a single valid fan. Valid fans may be closed around an interior vertex or open at a boundary while disconnected fans are non-manifold. (see Fig.~\ref{fig:topology_diagnostic} for illustration). 

    \item \emph{Global connectivity.} Two faces are considered adjacent if they share an edge. LCC faces and LCC area report the fractions of faces and surface area in the largest connected component (LCC), while Components counts the edge-connected components. We read component counts together with LCC fractions and visual inspection to account for differences in scene size and resolution. These three groups distinguish vertex sharing, local manifoldness, and fragmentation. App. Sec.~\ref{app:topology_metrics} gives the audit details.
\end{itemize}

\subsubsection{Training and Inference Efficiency}

We audit computational overhead across training and inference. During training, we track peak VRAM (MiB) and optimization time (min). During rendering, we report end-to-end throughput (FPS).

\section{Benchmark Evaluation}
\label{sec:experiments}

\subsection{Experiments Setup}
\label{sec:experiments_setup}

\textbf{Evaluated methods.} We evaluate four representative triangle- and mesh-based neural rendering methods, covering different designs in geometry connectivity, opacity, and appearance (Tab.~\ref{tab:feature_mapping}):

\begin{itemize}
	\item \textbf{2DTS} \citep{2dts} replaces Gaussian primitives with independent planar triangle facelets and renders them through differentiable alpha blending. Its opacity function is defined using barycentric coordinates, with a compactness parameter controlling the transition from diffuse splats toward sharper triangle surfaces.

	\item \textbf{Triangle-Splatting} \citep{trianglesplatting} directly optimizes independent triangles using a differentiable window function confined to each projected triangle. Learned opacity, view-dependent color, and depth-ordered alpha blending support image-based optimization, while adaptive densification and pruning refine the allocation of triangles.

	\item \textbf{DiffSoup} \citep{diffsoup} represents scenes with a compact triangle soup carrying neural texture features and binary opacity. Stochastic opacity masking makes binary opacity trainable within a depth-tested rasterization pipeline, while a neural color decoder maps texture features to rendered colors.

	\item \textbf{MeshSplatting} \citep{meshsplatting} constructs a shared-vertex mesh through restricted Delaunay triangulation and jointly refines geometry and per-vertex spherical-harmonic appearance. Opacity and window parameter schedules progressively guide optimization toward opaque triangles.

\end{itemize}

The first three methods produce \emph{triangle soups} without explicit shared-vertex connectivity, whereas MeshSplatting produces an \emph{indexed mesh}. We audit the structure of its exported mesh in Sec.~\ref{sec:mesh_splatting_analysis}. We additionally include 3DGS \citep{3dgs} and 2DGS \citep{2dgs} as Gaussian-based reference baselines.

\textbf{Datasets.} We evaluate on four complementary datasets spanning real-world scenes and object-centric captures:
\begin{itemize}
	\item \textbf{Mip-NeRF~360} \citep{mipnerf360} comprises nine real scenes, with surrounding camera trajectories, cluttered geometry, and large depth variations that challenge reconstruction of both foreground and distant content.
	\item For \textbf{Tanks and Temples} \citep{tanksandtemples}, we use the two outdoor scenes \emph{train} and \emph{truck}, featuring large real-world objects, complex backgrounds, and occlusions.
	\item \textbf{DTU} \citep{dtu} provides laboratory captures of varied objects from 49 or 64 calibrated camera positions per scene, together with structured-light reference scans, enabling geometric evaluation.
	\item \textbf{NeRF-Synthetic} \citep{nerf} contains eight object-centric scenes with detailed geometry and realistic non-Lambertian materials, complementary to real captures.
\end{itemize}

\textbf{Implementation details.} We follow the original training pipelines while applying the shared data and evaluation protocols. On Mip-NeRF~360, all methods train and evaluate on the downsampled images released with the dataset ($4\times$ outdoors, $2\times$ indoors), as recommended by NerfBaselines~\citep{nerfbaselines}, since on-the-fly downsampling can shift PSNR by 0.18$\sim$0.30~dB and change method rankings; DiffSoup follows its official implementation and uses the $4\times$ images indoors. All training and evaluation use a single NVIDIA GeForce RTX~4090 GPU with 24\,GB memory.

\begin{figure}[!t]
\centering
\includegraphics[width=\textwidth]{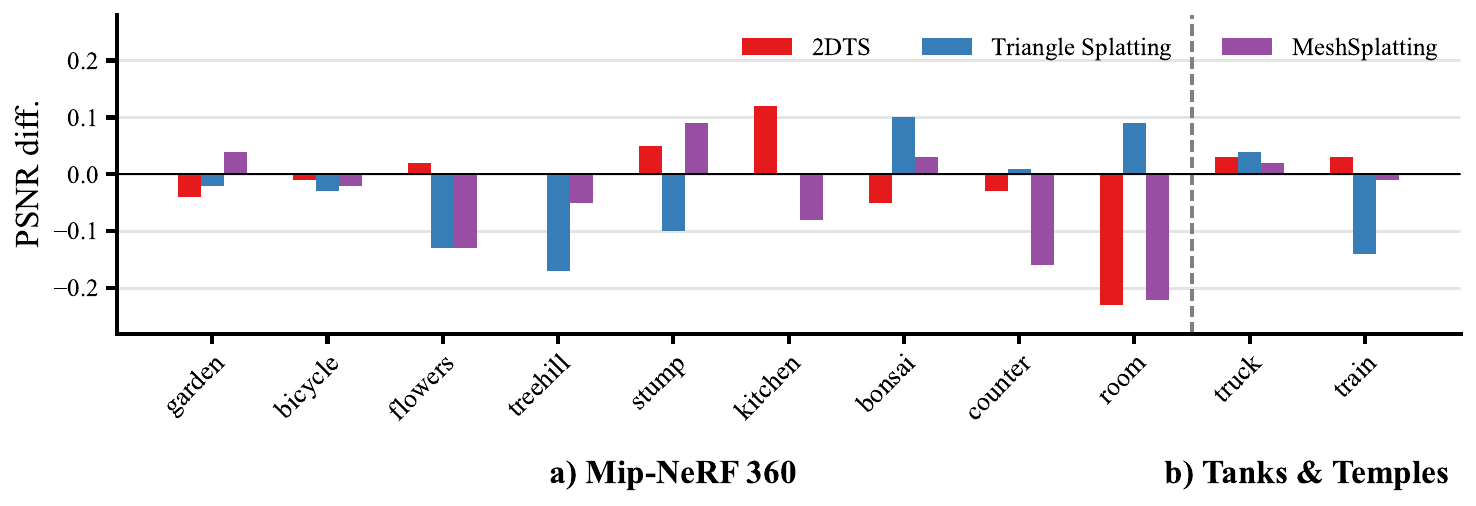}
\caption{Per-scene PSNR reproduction differences ($\mathrm{PSNR}_{\mathrm{MeshSuite}}-\mathrm{PSNR}_{\mathrm{paper}}$) on (a) Mip-NeRF~360 and (b) Tanks and Temples datasets. The largest relative deviation is below 0.8\%.
}
\label{fig:psnr_repro_comp_mipnerf360}
\end{figure}

\textbf{Throughput measurement.} We report only end-to-end throughput (FPS). Engine FPS is measured identically for all methods and both engine settings in a standalone Unity Player under Vulkan, as 1000 divided by the median wall-clock interval between consecutive frames, excluding image readback and file I/O; all engine frames are rendered with 2$\times$ MSAA. De facto renderer FPS is measured in each method's original pipeline. Dataset-level FPS is the mean of per-scene FPS.

\subsection{Standardized de facto renderer benchmark}
\label{sec:benchmark_reproduction}

We validate our reproductions against published results and public codebases. Across all 27 Mip-NeRF~360 method--scene pairs with published per-scene results, the mean absolute PSNR difference is only 0.075~dB and the maximum relative deviation is below 0.8\%. Fig.~\ref{fig:psnr_repro_comp_mipnerf360} shows this close agreement in image quality. These checks use each method's original image pipeline; all benchmark results use the shared downsampled images (for 2DTS on Mip-NeRF~360, 28.16 vs.\ 27.25~dB; App. Sec.~\ref{app:reproduction}). App. Tab.~\ref{tab:reproducibility_comparison} provides dataset-level comparisons against paper values and public-code results.

\begin{table}[tbp]
\centering
\footnotesize
\resizebox{\linewidth}{!}{%
\begin{tabular}{llcccccc@{}}
\toprule
\textbf{Dataset} & \textbf{Method} & \textbf{PSNR} $\uparrow$ & \textbf{SSIM} $\uparrow$ & \textbf{LPIPS} $\downarrow$ & \textbf{FPS} $\uparrow$ & \textbf{Peak mem.} $\downarrow$ & \textbf{Train time} $\downarrow$ \\
\midrule
\multirow{4}{*}{Mip-NeRF~360} & 2DTS & \textbf{27.25} & \textbf{0.812} & 0.245 & 70 & \textbf{10069} & 46.1 \\
& Triangle-Splatting & 27.13 & \textbf{0.812} & \textbf{0.227} & 94 & 18901 & 40.5 \\
& MeshSplatting & 24.72 & 0.729 & 0.365 & 38 & 23134 & 41.7 \\
& DiffSoup & 23.54 & 0.689 & 0.322 & \textbf{168} & 23484 & \textbf{16.8}\\
\midrule
\multirow{4}{*}{Tanks and Temples} & 2DTS & \textbf{23.42} & 0.853 & 0.203 & 101 & 16061 & 25.9 \\
& Triangle-Splatting & 23.08 & \textbf{0.856} & \textbf{0.171} & 159 & \textbf{11157} & 24.5 \\
& MeshSplatting & 20.53 & 0.756 & 0.319 & 66 & 21321 & 24.1 \\
& DiffSoup & 17.45 & 0.645 & 0.387 & \textbf{226} & 23890 & \textbf{15.4} \\
\midrule
\multirow{4}{*}{NeRF-Synthetic} & 2DTS & \textbf{33.65} & \textbf{0.970} & \textbf{0.034} & 126 & \textbf{3172} & 16.3 \\
& Triangle-Splatting & 29.87 & 0.956 & 0.055 & 243 & 12257 & 19.1 \\
& MeshSplatting & 29.50 & 0.945 & 0.071 & 86 & 23386 & 13.4 \\
& DiffSoup & 28.65 & 0.910 & 0.076 & \textbf{371} & 16334 & \textbf{12.2} \\
\bottomrule
\end{tabular}}
\caption{Standardized de facto renderer results for NVS quality and computational cost, averaged over scenes in each dataset.}
\label{tab:native_summary}
\end{table}

\begin{table}[tbp]
\centering
\footnotesize
\setlength{\tabcolsep}{1.5pt}
\begin{tabular*}{\linewidth}{@{\extracolsep{\fill}}l*{16}{c}@{}}
\toprule
\textbf{Method} & \textbf{24} & \textbf{37} & \textbf{40} & \textbf{55} & \textbf{63} & \textbf{65} & \textbf{69} & \textbf{83} & \textbf{97} & \textbf{105} & \textbf{106} & \textbf{110} & \textbf{114} & \textbf{118} & \textbf{122} & \textbf{Mean} \\
\midrule
2DTS & \textbf{0.49} & \textbf{0.61} & \textbf{0.37} & \textbf{0.44} & \textbf{0.73} & \textbf{0.79} & \textbf{0.64} & \textbf{0.87} & 1.23 & \textbf{0.42} & \textbf{0.54} & 1.00 & \textbf{0.45} & \textbf{0.46} & \textbf{0.41} & \textbf{0.63} \\
Triangle-Splatting & 1.08 & 1.09 & 1.05 & 0.73 & 1.86 & 1.54 & 1.20 & 1.58 & 1.91 & 1.02 & 0.85 & 1.22 & 0.96 & 1.08 & 1.81 & 1.26 \\
MeshSplatting & 0.77 & 0.70 & 0.77 & 0.66 & 0.87 & 1.09 & 0.78 & 1.02 & \textbf{1.20} & 0.56 & 0.77 & \textbf{0.86} & 0.71 & 0.74 & 0.69 & 0.81 \\
DiffSoup & 3.97 & 3.98 & 3.94 & 3.63 & 4.22 & 3.70 & 3.36 & 4.34 & 4.39 & 3.85 & 3.58 & 5.30 & 4.05 & 3.43 & 3.50 & 3.95 \\
\bottomrule
\end{tabular*}
\caption{Per-scene Chamfer distance on DTU \citep{dtu}.}
\label{tab:dtu_chamfer_per_scene}
\end{table}

Tab.~\ref{tab:native_summary} reports NVS quality and computational cost for the de facto renderers on Mip-NeRF~360 \citep{mipnerf360}, Tanks and Temples \citep{tanksandtemples}, and NeRF-Synthetic \citep{nerf}. 2DTS achieves the highest PSNR on Mip-NeRF~360 and NeRF-Synthetic (27.25~dB and 33.65~dB), while using the least peak training memory on NeRF-Synthetic (3172~MiB). Triangle-Splatting achieves the lowest LPIPS on Mip-NeRF~360 and Tanks and Temples (0.227 and 0.171). DiffSoup renders and trains fastest on all three datasets (up to 371~FPS and 12.2$\sim$16.8~min), but has lower image quality (17.45~dB on Tanks and Temples). MeshSplatting uses shared-vertex connectivity, with lower PSNR on Mip-NeRF~360 (24.72~dB). Thus, \textit{no} representation dominates fidelity, throughput, memory, and training time simultaneously.

\textbf{Surface geometry on DTU.}
Tab.~\ref{tab:dtu_chamfer_per_scene} complements the NVS quality and computational cost results in Tab.~\ref{tab:native_summary} with surface accuracy against DTU reference scans. 2DTS achieves the lowest mean CD (0.63~mm), followed by MeshSplatting (0.81~mm), Triangle-Splatting (1.26~mm), and DiffSoup (3.95~mm). 2DTS has the lowest CD on 13 of the 15 scenes, while MeshSplatting performs best on \emph{scan97} and \emph{scan110} (1.20~mm and 0.86~mm). These results supplement image-quality and efficiency measurements with direct evaluation of 3D surface accuracy before graphics deployment.

\subsection{Evaluating Graphics Engine Deployment}
\label{sec:renderer_deployment}

Tab.~\ref{tab:mip360_renderer_comparison} compares quality and throughput across the three settings on Mip-NeRF~360 and Tanks and Temples; App. Fig.~\ref{fig:deployment_gap} summarizes the PSNR trajectories of the four triangle- and mesh-based methods. App. Tabs.~\ref{tab:mip360_per_scene_quality} and~\ref{tab:tnt_renderer_comparison} retain their per-scene Mip-NeRF~360 and Tanks and Temples results. Under our deployment protocol, 3DGS \citep{3dgs} is unsupported in both engine settings; 2DGS \citep{2dgs} lacks custom engine deployment because TSDF meshing discards appearance variables used by the de facto renderer (though its mesh runs under standard engine deployment); and DiffSoup, whose neural feature decoder has no counterpart in the opaque vertex-color pipeline, enters standard engine deployment through its baked RGBA texture atlas with alpha cutout. App. Sec.~\ref{app:additional_results} additionally reports NeRF-Synthetic deployment, and App. Sec.~\ref{app:training_variants} evaluates engine-oriented training variants of 2DTS and Triangle-Splatting.

\textbf{Deployment quality.} The evaluated triangle- and mesh-based methods rely on specialized appearance models or compositing operations that limit their direct compatibility with conventional opaque mesh pipelines. Our standard engine deployment tests this compatibility using static vertex colors and hardware Z-buffering. Methods that depend on rendering operations beyond this pipeline can lose substantial image quality when adapted to it.

\begin{table}[t]
\centering
\textbf{(a) Mip-NeRF~360}\par\smallskip
\footnotesize
\setlength{\tabcolsep}{1.5pt}
\begin{tabular*}{\linewidth}{@{\extracolsep{\fill}}llcccccc@{}}
\toprule
 &  & \multicolumn{2}{c}{\textbf{Reference baselines}} & \multicolumn{4}{c}{\textbf{Triangle-/mesh-based methods}} \\
\cmidrule(lr){3-4}\cmidrule(l){5-8}
\textbf{Setting} & \textbf{Metric} & \textbf{3DGS} & \textbf{2DGS} & \textbf{2DTS} & \textbf{Triangle-Splatting} & \textbf{DiffSoup} & \textbf{MeshSplatting} \\
\midrule
\multirow{4}{*}{\shortstack[l]{\textbf{De facto renderer}}}
 & PSNR $\uparrow$ & 27.21 & 26.79 & \textbf{27.25} & 27.13 & 23.54 & 24.72 \\
 & SSIM $\uparrow$ & 0.815 & 0.796 & \textbf{0.812} & \textbf{0.812} & 0.689 & 0.729 \\
 & LPIPS $\downarrow$ & 0.214 & 0.252 & 0.245 & \textbf{0.227} & 0.322 & 0.365 \\
 & FPS $\uparrow$ & 109 & 38 & 70 & 94 & \textbf{168} & 38 \\
\midrule
\multirow{4}{*}{\shortstack[l]{\textbf{Standard engine}\\\textbf{deployment}}}
 & PSNR $\uparrow$ & \multirow{4}{*}{\textsc{N/A}} & 12.70 & 11.67 & 14.61 & \textbf{22.20} & 20.96 \\
 & SSIM $\uparrow$ &  & 0.311 & 0.299 & 0.252 & \textbf{0.649} & 0.572 \\
 & LPIPS $\downarrow$ &  & 0.594 & 0.696 & 0.662 & \textbf{0.366} & 0.484 \\
 & FPS $\uparrow$ &  & -- & 1555 & 1380 & \textbf{1651} & 909 \\
\midrule
\multirow{4}{*}{\shortstack[l]{\textbf{Custom engine}\\\textbf{deployment}}}
 & PSNR $\uparrow$ & \multirow{4}{*}{\textsc{N/A}} & \multirow{4}{*}{\textsc{N/A}} & \textbf{27.25} & 27.11 & 23.49 & 24.71 \\
 & SSIM $\uparrow$ &  &  & \textbf{0.813} & 0.812 & 0.679 & 0.729 \\
 & LPIPS $\downarrow$ &  &  & 0.244 & \textbf{0.227} & 0.325 & 0.366 \\
 & FPS $\uparrow$ &  &  & 230 & 291 & \textbf{1661} & 156 \\
\bottomrule
\end{tabular*}
\par\medskip
\textbf{(b) Tanks and Temples}\par\smallskip
\begin{tabular*}{\linewidth}{@{\extracolsep{\fill}}llcccccc@{}}
\toprule
 &  & \multicolumn{2}{c}{\textbf{Reference baselines}} & \multicolumn{4}{c}{\textbf{Triangle-/mesh-based methods}} \\
\cmidrule(lr){3-4}\cmidrule(l){5-8}
\textbf{Setting} & \textbf{Metric} & \textbf{3DGS} & \textbf{2DGS} & \textbf{2DTS} & \textbf{Triangle-Splatting} & \textbf{DiffSoup} & \textbf{MeshSplatting} \\
\midrule
\multirow{4}{*}{\shortstack[l]{\textbf{De facto renderer}}}
 & PSNR $\uparrow$ & 23.74 & 23.21 & \textbf{23.42} & 23.08 & 17.45 & 20.53 \\
 & SSIM $\uparrow$ & 0.853 & 0.832 & 0.853 & \textbf{0.856} & 0.645 & 0.756 \\
 & LPIPS $\downarrow$ & 0.169 & 0.211 & 0.203 & \textbf{0.171} & 0.387 & 0.319 \\
 & FPS $\uparrow$ & 129 & 92 & 101 & 159 & \textbf{226} & 66 \\
\midrule
\multirow{4}{*}{\shortstack[l]{\textbf{Standard engine}\\\textbf{deployment}}}
 & PSNR $\uparrow$ & \multirow{4}{*}{\textsc{N/A}} & 16.10 & 11.84 & 10.38 & \textbf{16.85} & 15.76 \\
 & SSIM $\uparrow$ &  & 0.525 & 0.337 & 0.299 & \textbf{0.625} & 0.599 \\
 & LPIPS $\downarrow$ &  & 0.515 & 0.655 & 0.684 & \textbf{0.419} & 0.455 \\
 & FPS $\uparrow$ &  & -- & 1564 & 1750 & \textbf{1777} & 1382 \\
\midrule
\multirow{4}{*}{\shortstack[l]{\textbf{Custom engine}\\\textbf{deployment}}}
 & PSNR $\uparrow$ & \multirow{4}{*}{\textsc{N/A}} & \multirow{4}{*}{\textsc{N/A}} & \textbf{23.43} & 23.07 & 17.44 & 20.54 \\
 & SSIM $\uparrow$ &  &  & 0.852 & \textbf{0.855} & 0.646 & 0.758 \\
 & LPIPS $\downarrow$ &  &  & 0.201 & \textbf{0.172} & 0.393 & 0.320 \\
 & FPS $\uparrow$ &  &  & 447 & 525 & \textbf{2114} & 296 \\
\bottomrule
\end{tabular*}
\caption{Mean NVS quality and throughput across the three evaluation settings
on (a) Mip-NeRF~360 and (b) Tanks and Temples. FPS is end-to-end throughput.}
\label{tab:mip360_renderer_comparison}
\end{table}

\begin{figure}[tb]
\centering
\includegraphics[width=\linewidth]{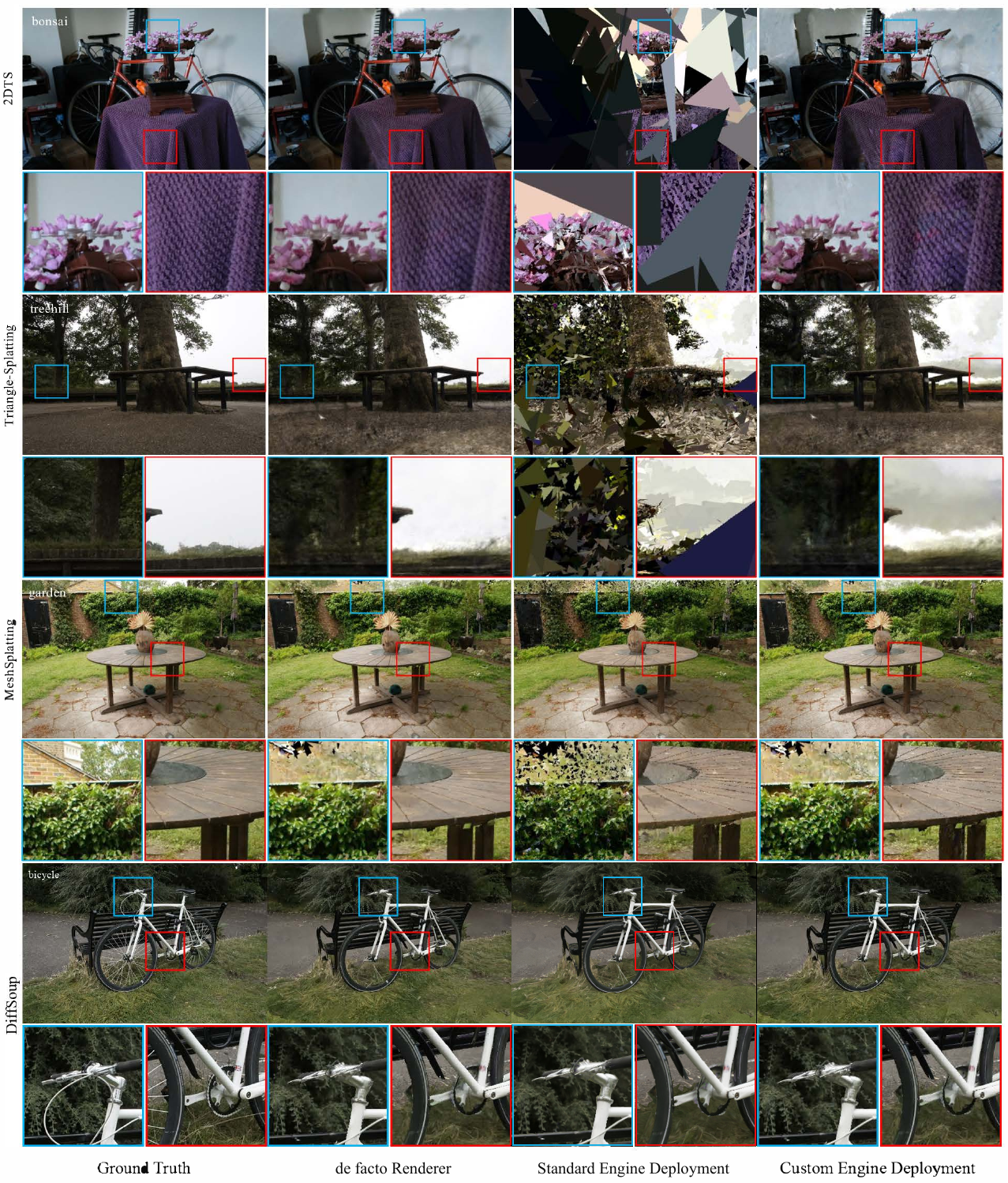}
\caption{Qualitative NVS comparisons of 2DTS \citep{2dts}, Triangle-Splatting \citep{trianglesplatting}, MeshSplatting \citep{meshsplatting}, and DiffSoup \citep{diffsoup} on Mip-NeRF~360. Columns show ground truth, de facto renderer, standard engine deployment, and custom engine deployment. DiffSoup's standard engine deployment renders its baked texture atlas.}
\label{fig:qualitative_deployment_2dts}
\end{figure}

On Mip-NeRF~360, adapting 2DTS, Triangle-Splatting, and MeshSplatting to standard engine deployment lowers mean PSNR by 15.58~dB, 12.52~dB, and 3.76~dB, respectively, relative to the de facto renderer (Tab.~\ref{tab:mip360_renderer_comparison}). MeshSplatting has the smallest overall deployment gap ($\Delta Q_m^{\mathrm{deploy}}$) among these three methods. The rendered examples illustrate the visual losses: 2DTS produces large opaque shards on \emph{bonsai}, and Triangle-Splatting exhibits a similar failure on \emph{treehill} (Fig.~\ref{fig:qualitative_deployment_2dts}).

We implement method-specific Unity renderers for custom engine deployment to retain each method's appearance and compositing features. On Mip-NeRF~360, these implementations recover 3.75$\sim$15.58~dB PSNR over standard engine deployment across 2DTS, Triangle-Splatting, and MeshSplatting, corresponding to their portability gaps ($\Delta Q_m^{\mathrm{port}}$). The remaining adaptation gaps ($\Delta Q_m^{\mathrm{adapt}}$) to the de facto renderer are at most 0.02~dB for these three methods on both Mip-NeRF~360 and Tanks and Temples, because the custom renderers follow the source kernels (App. Sec.~\ref{app:custom_renderers}). App. Fig.~\ref{fig:deployment_gap} shows these changes across the three settings.

For DiffSoup, our custom engine deployment retains the color decoder required by its neural texture features and achieves a mean PSNR of 23.49~dB on Mip-NeRF~360, compared with 23.54~dB for the de facto renderer (Tab. \ref{tab:mip360_renderer_comparison}). Its standard engine deployment replaces the decoder with the baked texture atlas and reaches 22.20~dB, a deployment gap of only 1.34~dB (Fig.~\ref{fig:qualitative_deployment_2dts}, bottom row). Supplementary videos compare the rendering settings.

As a complementary baseline, 2DGS \citep{2dgs} illustrates a different deployment trade-off from DiffSoup. On Mip-NeRF~360, its de facto renderer achieves 26.79~dB PSNR, whereas the extracted TSDF mesh reaches only 12.70~dB PSNR under standard engine deployment (Tab.~\ref{tab:mip360_renderer_comparison}). Custom engine deployment is unavailable under our deployment protocol because the exported mesh does not retain the Gaussian splatting parameters and view-dependent appearance of the de facto renderer. App. Sec.~\ref{app:2dgs_qualitative} and Fig.~\ref{fig:qualitative_deployment_2dgs} provide comparisons on the \emph{counter} scene, showing softened specular reflections and blurred object boundaries after mesh conversion.

\textbf{Cross-dataset deployment behavior.} The leading method depends on the deployment setting and the metric, with the same pattern on both datasets. Under custom engine deployment, 2DTS has the highest mean PSNR (27.25~dB on Mip-NeRF~360 and 23.43~dB on Tanks and Temples), exceeding Triangle-Splatting by 0.14~dB and 0.36~dB, while Triangle-Splatting has the lowest LPIPS (0.227 and 0.172). DiffSoup has the highest custom engine deployment throughput on both datasets, reaching 1661~FPS on Mip-NeRF~360 and 2114~FPS on Tanks and Temples. Under standard engine deployment, DiffSoup's baked texture atlas leads in all three image metrics on both datasets (22.20~dB and 16.85~dB PSNR). Among the three vertex-colored methods with standard NVS training, MeshSplatting achieves the best standard engine deployment quality on both datasets; on Tanks and Temples, it reaches 15.76~dB PSNR, 0.599 SSIM, and 0.455 LPIPS. On Tanks and Temples, custom engine deployment improves all three image metrics for every method; it reduces throughput by 3.3$\sim$4.7$\times$ for 2DTS, Triangle-Splatting, and MeshSplatting, but not for DiffSoup (2114 vs.\ 1777~FPS).

\textbf{Engine-oriented training.} These rankings apply to standard NVS training. With engine-oriented training, \textit{i.e.}, opaque 2DTS and Triangle-Splatting's official game-engine configuration (App. Sec.~\ref{app:training_variants}), standard engine deployment PSNR on Mip-NeRF~360 rises from 11.67 to 20.30~dB and from 14.61 to 18.31~dB; on Tanks and Temples, 2DTS (opaque) reaches 17.69~dB, above MeshSplatting (15.76~dB) and baked DiffSoup (16.85~dB). Their custom engine deployment slightly exceeds their own de facto renderers (by 0.13$\sim$0.34~dB), but their de facto renderer PSNR is 3.86$\sim$5.41~dB lower than with NVS training, and standard engine deployment still trails it by 1.87$\sim$3.41~dB. Standard engine deployment thus degrades every configuration, and its quality depends on opaque training rather than on the method family.

\textbf{Quality-efficiency trade-off.} Custom engine deployment recovers quality lost when adapting the methods to the standard opaque mesh pipeline, but reduces throughput for the alpha-blended methods (Tab.~\ref{tab:mip360_renderer_comparison}). On Mip-NeRF~360, Triangle-Splatting gains 12.50~dB PSNR over standard engine deployment with a 4.7$\times$ reduction in throughput (291 vs.\ 1380~FPS). For 2DTS, the gain is 15.58~dB with a 6.8$\times$ reduction (230 vs.\ 1555~FPS). Across all four methods on this dataset, custom engine deployment reaches 156$\sim$1661~FPS, while standard engine deployment reaches 909$\sim$1651~FPS. The corresponding reductions in throughput are 4.7$\sim$6.8$\times$ for 2DTS, Triangle-Splatting, and MeshSplatting, whereas DiffSoup's throughput is unchanged (1661 vs.\ 1651~FPS). De facto renderer FPS is measured in each method's original pipeline rather than in the engine and is shown for reference.

\begin{table}[t]
\centering
\small
\begin{tabular*}{\linewidth}{@{\extracolsep{\fill}}lcccccccc@{}}
\toprule
\multirow{2}{*}{\textbf{Dataset}} &
\multicolumn{2}{c}{\textbf{Local reuse}} &
\multicolumn{3}{c}{\textbf{Local structure}} &
\multicolumn{3}{c}{\textbf{Global connectivity}} \\
\cmidrule(lr){2-3}\cmidrule(lr){4-6}\cmidrule(l){7-9}
& \textbf{V/F} & \textbf{Val.}
& \shortstack{\textbf{Bnd.}\\\textbf{edges}}
& \shortstack{\textbf{NM}\\\textbf{edges} $\downarrow$}
& \shortstack{\textbf{NM}\\\textbf{verts.} $\downarrow$}
& \shortstack{\textbf{LCC}\\\textbf{faces} $\uparrow$}
& \shortstack{\textbf{LCC}\\\textbf{area} $\uparrow$}
& \shortstack{\textbf{Comp.}\\\textbf{(K)}} \\
\midrule
\multicolumn{9}{c}{\textit{Large, unbounded scenes}} \\
\midrule
Mip-NeRF~360 & 0.523 & 6.49 & 0.454 & 0.178 & 0.652 & 0.503 & 0.174 & 286.6 \\
Tanks and Temples & 0.516 & 6.49 & 0.461 & 0.182 & 0.665 & 0.668 & 0.176 & 171.7 \\
\midrule
\multicolumn{9}{c}{\textit{Object-centric scenes}} \\
\midrule
DTU & 0.428 & 7.36 & 0.412 & 0.222 & 0.727 & 0.850 & 0.806 & 15.6 \\
NeRF-Synthetic & 0.339 & 8.63 & 0.372 & 0.284 & 0.873 & 0.674 & 0.713 & 11.3 \\
\bottomrule
\end{tabular*}
\caption{Mesh diagnostics for MeshSplatting \citep{meshsplatting}, averaged over all scenes in each dataset. Val.: valence; Bnd.: boundary; NM: non-manifold; Comp.: edge-connected components.}
\label{tab:mesh_connectivity}
\end{table}

\subsection{Mesh Structural Results}
\label{sec:mesh_splatting_analysis}
We audit MeshSplatting's exported indexed meshes across all four datasets using the three metric groups defined in Sec.~\ref{sec:deployment_gap}. Tab.~\ref{tab:mesh_connectivity} reports the mean of each metric over all scenes in each dataset. We consider the large, unbounded scenes of Mip-NeRF~360 and Tanks and Temples separately from the smaller, object-centric scenes of DTU and NeRF-Synthetic. App. Sec.~\ref{app:topology_metrics} provides per-scene results and audit details. We highlight the following results:

\begin{itemize}
    \item \emph{Local reuse.} Mip-NeRF~360 and Tanks and Temples exhibit similar vertex sharing, with mean $|\mathbf{V}|/|\mathbf{F}|$ ratios of 0.523 and 0.516 and mean valence of 6.49 on both datasets. The object-centric datasets show greater reuse: DTU and NeRF-Synthetic have lower $|\mathbf{V}|/|\mathbf{F}|$ ratios of 0.428 and 0.339, with higher valence of 7.36 and 8.63, respectively. Vertex sharing is evident in both scene groups, but its extent differs.
    
    \item \emph{Local structure.} Vertex sharing does not ensure manifoldness. On Mip-NeRF~360 and Tanks and Temples, approximately 18\% of edges and 65$\sim$66\% of vertices are non-manifold. On DTU and NeRF-Synthetic, the non-manifold-edge ratios rise to 22\% and 28\%, and the non-manifold-vertex ratios to 73\% and 87\%, respectively. Thus, greater reuse on the object-centric datasets coexists with more non-manifold neighborhoods. Boundary edges account for 37$\sim$46\% of edges across the four datasets, indicating open surfaces while such boundaries can be expected in partially observed scenes. Non-manifold edges and vertices, in contrast, violate local manifold connectivity, and lower ratios indicate better local structural integrity.

    \item \emph{Global connectivity.} On Mip-NeRF~360 and Tanks and Temples, the largest connected component (LCC) contains 50.3\% and 66.8\% of faces but only 17.4\% and 17.6\% of surface area, respectively. Thus, a majority of faces in one component does not imply comparable surface-area coverage: over 82\% of area lies in other components on both large-scene datasets. DTU and NeRF-Synthetic show higher LCC area fractions of 80.6\% and 71.3\%, with LCC face fractions of 85.0\% and 67.4\%, respectively, but still leave 19.4\% and 28.7\% of area outside the LCC. Mean component counts are 286.6k and 171.7k on the large-scene datasets, compared with 15.6k and 11.3k on the object-centric datasets. These results reveal fragmentation in both scene groups, although absolute component counts depend on mesh size and do not alone provide a structural-quality ranking across datasets. On the two deployment datasets, MeshSplatting's leading standard engine deployment image quality among the three vertex-colored methods with NVS training (Tab.~\ref{tab:mip360_renderer_comparison}) coexists with these structural defects, showing that visual fidelity does not ensure structural integrity.
\end{itemize}

\section{Conclusion}
We introduced MeshSuite, a unified benchmark for triangle- and mesh-based neural rendering that connects evaluation with de facto renderers to graphics-engine deployment. MeshSuite combines standardized evaluation protocols, a unified asset interface, and method-specific Unity renderers to compare the de facto renderer, standard engine deployment, and custom engine deployment. It evaluates visual fidelity and computational cost alongside surface accuracy and mesh structural integrity.

Our experiments show that visual quality and method rankings depend on the deployment setting. Custom engine deployment matches the de facto renderers to within 0.05~dB PSNR but incurs higher runtime costs than standard engine deployment, which degrades every method and training configuration. Among the three vertex-colored methods, MeshSplatting achieves the highest image quality under standard engine deployment on Mip-NeRF~360 and Tanks and Temples, second only to DiffSoup's baked texture atlas, yet its exported meshes exhibit non-manifold structures and substantial fragmentation across both large, unbounded and object-centric scenes. These findings show that triangle rasterization and shared-vertex connectivity alone do not ensure practical graphics usability. Evaluation must jointly consider appearance preservation, rendering efficiency, and the structural requirements of downstream tasks.
Future work can extend the benchmark to additional graphics engines and directly assess exported meshes in tasks such as editing, remeshing, and collision handling.




\newpage
\bibliographystyle{iclr2027_conference}
\bibliography{iclr2027_conference}

@inproceedings{2dgs,
  author    = {Huang, Binbin and Yu, Zehao and Chen, Anpei and Geiger, Andreas and Gao, Shenghua},
  booktitle = {ACM SIGGRAPH 2024 Conference Papers},
  pages     = {1--11},
  publisher = {ACM},
  title     = {{2D} Gaussian Splatting for Geometrically Accurate Radiance Fields},
  year      = {2024}
}

@misc{2dts,
  author    = {Sheng, Kaifeng and Zhou, Zheng and Peng, Yingliang and Wang, Qianwei},
  archivePrefix = {arXiv},
  eprint    = {2506.18575},
  primaryClass = {cs.CV},
  title     = {{2D} Triangle Splatting for Direct Differentiable Mesh Training},
  year      = {2025}
}

@article{3dgs,
  author    = {Kerbl, Bernhard and Kopanas, Georgios and Leimk{\"u}hler, Thomas and Drettakis, George},
  journal   = {ACM Transactions on Graphics},
  number    = {4},
  pages     = {1--14},
  publisher = {ACM},
  title     = {{3D} Gaussian Splatting for Real-Time Radiance Field Rendering},
  volume    = {42},
  year      = {2023}
}

@inproceedings{3dgs-mcmc,
  author    = {Kheradmand, Shakiba and Rebain, Daniel and Sharma, Gopal and Sun, Weiwei and Tseng, Yang-Che and Isack, Hossam and Kar, Abhishek and Tagliasacchi, Andrea and Yi, Kwang Moo},
  booktitle = {Advances in Neural Information Processing Systems},
  note      = {Spotlight Presentation},
  title     = {3D Gaussian Splatting as Markov Chain Monte Carlo},
  year      = {2024}
}

@misc{blender,
  author = {{Blender Foundation}},
  howpublished = {Online},
  title = {Blender},
  url = {https://www.blender.org/},
  year = {2026}
}

@article{broxton2020immersive,
  author = {Broxton, Michael and Flynn, John and Overbeck, Ryan and Erickson, Daniel and Hedman, Peter and DuVall, Matthew and Dourgarian, Jason and Busch, Jay and Whalen, Matt and Debevec, Paul},
  title = {Immersive Light Field Video with a Layered Mesh Representation},
  journal = {ACM Transactions on Graphics},
  volume = {39},
  number = {4},
  year = {2020}
}

@inproceedings{diffsoup,
  author    = {Tojo, Kenji and Bickel, Bernd and Umetani, Nobuyuki},
  booktitle = {Proceedings of the IEEE/CVF Conference on Computer Vision and Pattern Recognition},
  pages     = {8353--8363},
  title     = {DiffSoup: Direct Differentiable Rasterization of Triangle Soup for Extreme Radiance Field Simplification},
  year      = {2026}
}

@inproceedings{dtu,
  author    = {Jensen, Rasmus and Dahl, Anders and Vogiatzis, George and Tola, Engin and Aan{\ae}s, Henrik},
  booktitle = {Proceedings of the IEEE conference on computer vision and pattern recognition},
  pages     = {406--413},
  title     = {Large scale multi-view stereopsis evaluation},
  year      = {2014}
}

@article{instantngp,
  author    = {M{\"u}ller, Thomas and Evans, Alex and Schied, Christoph and Keller, Alexander},
  journal   = {ACM Transactions on Graphics},
  number    = {4},
  pages     = {1--15},
  publisher = {ACM},
  title     = {Instant Neural Graphics Primitives with a Multiresolution Hash Encoding},
  volume    = {41},
  year      = {2022}
}

@inproceedings{kilonerf,
  author    = {Reiser, Christian and Peng, Songyou and Liao, Yiyi and Geiger, Andreas},
  booktitle = {Proceedings of the IEEE/CVF International Conference on Computer Vision},
  pages     = {14335--14345},
  title     = {{KiloNeRF}: Speeding Up Neural Radiance Fields With Thousands of Tiny {MLP}s},
  year      = {2024}
}

@inproceedings{liang2024perceptual,
  author       = {Liang, Hanxue and Wu, Tianhao and Hanji, Param and Banterle, Francesco and Gao, Hongyun and Mantiuk, Rafal and {\"O}ztireli, Cengiz},
  booktitle    = {Computer Graphics Forum},
  organization = {Wiley Online Library},
  pages        = {e15036},
  title        = {Perceptual Quality Assessment of NeRF and Neural View Synthesis Methods for Front-Facing Views},
  volume       = {43},
  year         = {2024}
}

@inproceedings{meshsplatting,
  author    = {Held, Jan and Son, Sanghyun and Vandeghen, Renaud and Rebain, Daniel and Gadelha, Matheus and Zhou, Yi and Cioppa, Anthony and Lin, Ming C. and Van Droogenbroeck, Marc and Tagliasacchi, Andrea},
  booktitle = {Proceedings of the IEEE/CVF Conference on Computer Vision and Pattern Recognition},
  pages     = {7320--7329},
  title     = {{MeshSplatting}: Differentiable Rendering with Opaque Meshes},
  year      = {2026}
}

@inproceedings{mipnerf,
  author    = {Barron, Jonathan T. and Mildenhall, Ben and Tancik, Matthew and Hedman, Peter and Martin-Brualla, Ricardo and Srinivasan, Pratul P.},
  booktitle = {Proceedings of the IEEE/CVF International Conference on Computer Vision},
  pages     = {5855--5864},
  title     = {{Mip-NeRF}: A Multiscale Representation for Anti-Aliasing Neural Radiance Fields},
  year      = {2021}
}

@inproceedings{mipnerf360,
  author    = {Barron, Jonathan T. and Mildenhall, Ben and Verbin, Dor and Srinivasan, Pratul P. and Hedman, Peter},
  booktitle = {Proceedings of the IEEE/CVF Conference on Computer Vision and Pattern Recognition},
  pages     = {5470--5479},
  title     = {{Mip-NeRF 360}: Unbounded Anti-Aliased Neural Radiance Fields},
  year      = {2022}
}

@inproceedings{nerf,
  author    = {Mildenhall, Ben and Srinivasan, Pratul P. and Tancik, Matthew and Barron, Jonathan T. and Ramamoorthi, Ravi and Ng, Ren},
  booktitle = {Proceedings of the European Conference on Computer Vision},
  pages     = {405--421},
  title     = {{NeRF}: Representing Scenes as Neural Radiance Fields for View Synthesis},
  year      = {2020}
}

@inproceedings{nice_slam,
  author = {Zhu, Zihan and Peng, Songyou and Larsson, Viktor and Xu, Weiwei and Bao, Hujun and Cui, Zhaopeng and Oswald, Martin R. and Pollefeys, Marc},
  title = {{NICE-SLAM}: Neural Implicit Scalable Encoding for {SLAM}},
  booktitle = {Proceedings of the IEEE/CVF Conference on Computer Vision and Pattern Recognition},
  pages = {12786--12796},
  year = {2022}
}

@inproceedings{plenoctrees,
  author    = {Yu, Alex and Li, Ruilong and Tancik, Matthew and Li, Hao and Ng, Ren and Kanazawa, Angjoo},
  booktitle = {Proceedings of the IEEE/CVF International Conference on Computer Vision},
  pages     = {5752--5761},
  title     = {{PlenOctrees} for Real-Time Rendering of Neural Radiance Fields},
  year      = {2021}
}

@inproceedings{plenoxels,
  author    = {Fridovich-Keil, Sara and Yu, Alex and Tancik, Matthew and Chen, Qinhong and Recht, Benjamin and Kanazawa, Angjoo},
  booktitle = {Proceedings of the IEEE/CVF Conference on Computer Vision and Pattern Recognition},
  pages     = {5501--5510},
  title     = {{Plenoxels}: Radiance Fields Without Neural Networks},
  year      = {2022}
}

@inproceedings{sugar,
  author    = {Gu{\'e}don, Antoine and Lepetit, Vincent},
  booktitle = {Proceedings of the IEEE/CVF Conference on Computer Vision and Pattern Recognition},
  pages     = {5354--5365},
  title     = {{SuGaR}: Surface-Aligned Gaussian Splatting for Efficient {3D} Mesh Reconstruction and High-Quality Mesh Rendering},
  year      = {2024}
}

@article{tanksandtemples,
  author  = {Arno Knapitsch and Jaesik Park and Qian-Yi Zhou and Vladlen Koltun},
  journal = {ACM Transactions on Graphics},
  number  = {4},
  title   = {Tanks and Temples: Benchmarking Large-Scale Scene Reconstruction},
  volume  = {36},
  year    = {2017}
}

@inproceedings{tensorf,
  author    = {Chen, Anpei and Xu, Zexiang and Geiger, Andreas and Yu, Jingyi and Su, Hao},
  booktitle = {Computer Vision -- ECCV 2022},
  pages     = {333--350},
  publisher = {Springer},
  title     = {{TensoRF}: Tensorial Radiance Fields},
  year      = {2022}
}

@inproceedings{trianglesplatting,
  author    = {Held, Jan and Vandeghen, Renaud and Deliege, Adrien and Hamdi, Abdullah and Rebain, Daniel and Giancola, Silvio and Cioppa, Anthony and Vedaldi, Andrea and Ghanem, Bernard and Tagliasacchi, Andrea and Van Droogenbroeck, Marc},
  booktitle = {International Conference on 3D Vision},
  title     = {Triangle Splatting for Real-Time Radiance Field Rendering},
  year      = {2026}
}

@misc{unity,
  author = {{Unity Technologies}},
  howpublished = {Online},
  title = {Unity},
  url = {https://unity.com/},
  year = {2026}
}

@techreport{unity_photogrammetry,
  author = {Lachambre, S{\'e}bastien and Lagarde, S{\'e}bastien and Jover, Cyril},
  title = {Photogrammetry Workflow},
  institution = {Unity Technologies},
  year = {2017}
}

@inproceedings{vgg,
  author    = {Karen Simonyan and Andrew Zisserman},
  booktitle = {International Conference on Learning Representations},
  title     = {Very Deep Convolutional Networks for Large-Scale Image Recognition},
  year      = {2015}
}

@article{ssim,
  author  = {Wang, Zhou and Bovik, Alan C. and Sheikh, Hamid R. and Simoncelli, Eero P.},
  journal = {IEEE Transactions on Image Processing},
  number  = {4},
  pages   = {600--612},
  title   = {Image Quality Assessment: From Error Visibility to Structural Similarity},
  volume  = {13},
  year    = {2004}
}

@inproceedings{zipnerf,
  author  = {Barron, Jonathan T. and Mildenhall, Ben and Verbin, Dor and Srinivasan, Pratul P. and Hedman, Peter},
  booktitle = {Proceedings of the IEEE/CVF International Conference on Computer Vision},
  pages   = {19697--19705},
  title   = {Zip-NeRF: Anti-Aliased Grid-Based Neural Radiance Fields},
  year    = {2023}
}

@inproceedings{lpips,
  title={The unreasonable effectiveness of deep features as a perceptual metric},
  author={Zhang, Richard and Isola, Phillip and Efros, Alexei A and Shechtman, Eli and Wang, Oliver},
  booktitle={Proceedings of the IEEE conference on computer vision and pattern recognition},
  pages={586--595},
  year={2018}
}

@article{nerfbaselines,
  title={Nerfbaselines: Consistent and reproducible evaluation of novel view synthesis methods},
  author={Kulhanek, Jonas and Sattler, Torsten},
  journal={Advances in Neural Information Processing Systems},
  volume={38},
  pages={127378--127403},
  year={2025}
}

@inproceedings{splatwizard,
  title={Splatwizard: A Benchmark Toolkit for 3D Gaussian Splatting Compression},
  author={Liu, Xiang and Zhou, Yimin and Wang, Jinxiang and Huang, Yujun and Xie, Shuzhao and Qin, Shiyu and Hong, Mingyao and Li, Jiawei and Wang, Yaowei and Wang, Zhi and Xia, Shu-Tao and Chen, Bin},
  booktitle={Proceedings of the IEEE/CVF Conference on Computer Vision and Pattern Recognition Findings},
  pages={2261--2271},
  year={2026}
}

@inproceedings{mobilenerf,
  title={MobileNeRF: Exploiting the Polygon Rasterization Pipeline for Efficient Neural Field Rendering on Mobile Architectures},
  author={Zhiqin Chen and Thomas Funkhouser and Peter Hedman and Andrea Tagliasacchi},
  booktitle={The Conference on Computer Vision and Pattern Recognition},
  year={2023}
}

@inproceedings{bakedsdf,
  title={Bakedsdf: Meshing neural sdfs for real-time view synthesis},
  author={Yariv, Lior and Hedman, Peter and Reiser, Christian and Verbin, Dor and Srinivasan, Pratul P and Szeliski, Richard and Barron, Jonathan T and Mildenhall, Ben},
  booktitle={ACM SIGGRAPH 2023 conference proceedings},
  pages={1--9},
  year={2023}
}

@article{games,
  title={Games: Mesh-based adapting and modification of gaussian splatting},
  author={Waczy{\'n}ska, Joanna and Borycki, Piotr and Tadeja, S{\l}awomir and Tabor, Jacek and Spurek, Przemys{\l}aw},
  journal={arXiv preprint arXiv:2402.01459},
  year={2024}
}

@inproceedings{psnr,
  title={Image quality metrics: PSNR vs. SSIM},
  author={Hore, Alain and Ziou, Djemel},
  booktitle={2010 20th international conference on pattern recognition},
  pages={2366--2369},
  year={2010},
  organization={IEEE}
}

@inproceedings{nerfstudio,
  title={Nerfstudio: A Modular Framework for Neural Radiance Field Development},
  author={Tancik, Matthew and Weber, Ethan and Ng, Evonne and Li, Ruilong and Yi, Brent and Kerr, Justin and Wang, Terrance and Kristoffersen, Alexander and Austin, Jake and Salahi, Kamyar and others},
  booktitle={ACM SIGGRAPH 2023 Conference Proceedings},
  year={2023}
}

@article{gaustudio,
  title={GauStudio: A Modular Framework for 3D Gaussian Splatting and Beyond},
  author={Ye, Chongjie and Nie, Yinyu and Chang, Jiahao and Chen, Yuantao and Zhi, Yihao and Han, Xiaoguang},
  journal={arXiv preprint arXiv:2403.19632},
  year={2024}
}

@article{gsplat,
  title={gsplat: An Open-Source Library for Gaussian Splatting},
  author={Ye, Vickie and Li, Ruilong and Kerr, Justin and Turkulainen, Matias and Yi, Brent and Pan, Zhuoyang and Seiskari, Otto and Ye, Jianbo and Hu, Jeffrey and Tancik, Matthew and Kanazawa, Angjoo},
  journal={Journal of Machine Learning Research},
  volume={26},
  number={34},
  pages={1--17},
  year={2025}
}

@misc{nvsbench,
  title={{nvs-bench}: Novel-View Synthesis Benchmark},
  author={{nvs-bench contributors}},
  howpublished={\url{https://nvs-bench.github.io/}},
  note={Open-source benchmark project},
  year={2026}
}

@inproceedings{wang2024robustness,
  title={Benchmarking Robustness in Neural Radiance Fields},
  author={Wang, Chen and Wang, Angtian and Li, Junbo and Yuille, Alan and Xie, Cihang},
  booktitle={Proceedings of the IEEE/CVF Conference on Computer Vision and Pattern Recognition Workshops},
  pages={2926--2936},
  year={2024}
}

\newpage
\appendix
\section{Software Architecture and Implementation}
\label{app:software_architecture}

This section details the modular software architecture of \textit{MeshSuite}, expanding the overview in Fig.~\ref{fig:benchmark_protocol_teaser}. The infrastructure is built with decoupling and modularity as first principles, ensuring that model training, format translation, engine-native deployment, and multi-dimensional evaluation operate independently through clearly specified interfaces.

\textbf{Modular component design.} MeshSuite is organized into five cohesive subsystems:
(1)~\emph{Dataset Management}: Standardizes canonical camera frames, optical axes, field-of-view definitions, and resolution downsampling across datasets. For masked object captures (\textit{e.g.}, DTU), this layer automatically computes object alpha mattes and applies the white-background compositing convention specified in Eq.~\ref{eq:white_background}.
(2)~\emph{Algorithm Pool and Training}: Provides standardized training harnesses wrapping the official implementations of 2DTS \citep{2dts}, Triangle-Splatting \citep{trianglesplatting}, MeshSplatting \citep{meshsplatting}, and DiffSoup \citep{diffsoup}. It follows each method's original training pipeline while applying shared data and evaluation protocols under a common single-GPU hardware environment.
(3)~\emph{Universal Translation Layer}: The central translation bridge implementing method-specific adapters $\mathcal{E}_m$. It parses trained model parameters and extracts explicit geometry, opacity fields, and appearance representations into standardized intermediate assets $\mathcal{A}_m$.
(4)~\emph{Graphics Engine Deployment}: Native engine renderers implemented in Unity using C\# Scriptable Render Pipelines and custom HLSL shaders. This layer provides both standard engine deployment and custom engine deployment.
(5)~\emph{Evaluation and Audit Suite}: Automated evaluation scripts computing image fidelity metrics (PSNR, SSIM, LPIPS), end-to-end throughput, and mesh topology audits.

\textbf{Asset serialization and decoupled execution.} Under the Universal Translation Layer, trained models are converted into portable asset files prior to engine execution. Independent triangle soup models (2DTS, Triangle-Splatting, DiffSoup) are serialized into flat vertex arrays and attribute textures, whereas indexed meshes (MeshSplatting) are serialized with shared vertex coordinate tables, triangle face indices, and per-vertex SH coefficients. Crucially, the resulting asset bundles are completely self-contained and ingested directly by the graphics engine, decoupling real-time deployment from PyTorch, CUDA extension compilers, and Python runtime overhead. This architectural separation allows new neural rendering representations to be benchmarked by implementing a single translation adapter, and enables future deployment to other graphics engines (\textit{e.g.}, Unreal Engine, Blender) without modifying upstream training pipelines.

\subsection{Custom Engine Renderers}
\label{app:custom_renderers}
Each custom renderer follows its method's source kernel in the details that affect image quality.
\textbf{2DTS and Triangle-Splatting.} The shaders follow the source kernels, and custom engine deployment matches the de facto renderer PSNR to within 0.02~dB on Mip-NeRF~360, Tanks and Temples, and NeRF-Synthetic.
\textbf{MeshSplatting.} After training, MeshSplatting is effectively opaque (opacity floor 0.9999, $\sigma=10^{-4}$). Its source rasterizer shows the first triangle in centroid-depth order and renders at $4\times$ resolution before area downsampling. The shader reproduces both rules: every fragment takes its triangle's centroid depth for the hardware depth test, and the scene is rendered into a $4\times$ target resolved with a box filter. Replacing these rules with a per-pixel depth buffer or with 2$\times$ MSAA at the output resolution lowers PSNR on \emph{bicycle} by 0.69 and 0.57~dB (1.30~dB together); with both rules, custom engine deployment matches the de facto renderer to within 0.04~dB on every Mip-NeRF~360 and Tanks and Temples scene.
\textbf{DiffSoup.} Both renderers discard fragments whose texture alpha is below 0.5, keep the nearest remaining fragment, and decode its texture features with the same MLP. Two details differ: the shader interpolates texture coordinates in screen space rather than perspective-correctly, and the engine applies 2$\times$ MSAA. On Mip-NeRF~360, screen-space interpolation lowers PSNR by 0.15~dB on average and MSAA raises it by 0.11~dB; the loss is largest on finely textured surfaces (\emph{garden} and \emph{kitchen}, 0.31 and 0.25~dB).

\section{Reproducibility Comparisons and Analysis}
\label{app:reproduction}
Tab.~\ref{tab:reproducibility_comparison} complements the per-scene PSNR reproduction differences illustrated in Fig.~\ref{fig:psnr_repro_comp_mipnerf360} with comprehensive dataset-level evaluations comparing our reproduced metrics against published paper numbers and official public repositories. Sec.~\ref{sec:experiments_setup} outlines the primary experimental setup.

\textbf{Quantitative reproduction fidelity.} For 2DTS, Triangle-Splatting, and MeshSplatting, MeshSuite closely matches published dataset-level PSNR on Mip-NeRF~360 and Tanks and Temples. On Mip-NeRF~360, the mean absolute PSNR difference across their 27 scene-method pairs is only 0.075~dB, and the maximum relative deviation across these pairs is below 0.8\%\footnote{DiffSoup is omitted since it does not report per-scene metrics in its original paper.}. Specifically, 2DTS reproduces at 28.16~dB PSNR with its original image pipeline (vs.~28.18~dB in the paper and 28.51~dB in the public repository), Triangle-Splatting achieves 27.13~dB (vs.~27.17~dB reported and 27.20~dB in public code), and MeshSplatting achieves 24.72~dB (vs. 24.78~dB in both paper and code). For DiffSoup, where per-scene Mip-NeRF~360 results were not reported in the original paper, MeshSuite is closer to the public codebase in dataset-level PSNR (23.54~dB vs.~23.69~dB) than to the paper (24.76~dB); on Tanks and Temples, it achieves 17.45~dB vs.~18.38~dB in public code. LPIPS differences are larger for several methods, so close PSNR agreement does not imply equal agreement across all image metrics.

\textbf{Sources of discrepancies and impact of standardization.} Differences between published figures, public repositories, and MeshSuite may reflect preprocessing, camera conventions, background compositing, metric implementations, and numerical variation. We do not isolate their individual contributions:
(1)~\emph{Image Downsampling}: Image interpolation filters can affect reconstructed image values; MeshSuite uses the downsampled images released with each dataset for every method (Sec.~\ref{sec:experiments_setup}).
(2)~\emph{Camera Intrinsics and Principal Points}: Differences in principal point handling and coordinate conventions can affect alignment; MeshSuite enforces standardized projection matrices aligned with graphics conventions.
(3)~\emph{Background Compositing Conventions}: On masked datasets such as DTU, prior works alternated between evaluating against unmasked black backgrounds and masked white backgrounds; MeshSuite standardizes all evaluations to the alpha matte white-background protocol (Eq.~\ref{eq:white_background}).
(4)~\emph{Metric Implementations and Numerical Variation}: MeshSuite uses a common metric implementation, including VGG-based LPIPS (Sec.~\ref{sec:nvs_quality}), while parallel floating-point accumulation can also introduce numerical variation. The comparisons document agreement and remaining differences rather than establishing a single cause for them.

\textbf{2DTS on Mip-NeRF~360: original vs.\ shared images.} The 2DTS reproduction (28.16~dB) uses 2DTS's original data pipeline, which loads the full-resolution photographs and downsamples them on the fly with bilinear filtering, rounding the size down. All other 2DTS results in this paper come from models trained and evaluated on the released downsampled images used for every method. The two sets of targets differ in the downsampling filter and by one pixel per dimension (\textit{e.g.}, $1236\times821$ vs.\ $1237\times822$ on \emph{bicycle}), which also slightly changes the intrinsics. NerfBaselines~\citep{nerfbaselines} reports that on-the-fly downsampling raises Mip-NeRF~360 PSNR by 0.18$\sim$0.30~dB for seven methods and can change their ranking, and recommends the released images. For 2DTS, this difference accounts for the gap between 28.16~dB and 27.25~dB: the original-pipeline models score 27.31~dB against the shared images (lower on all 25 \emph{bicycle} views), and blurring the shared \emph{bicycle} targets with a 0.5-pixel Gaussian raises their score from 25.20 to 26.43~dB, close to the 26.30~dB measured against the original-pipeline targets. Under the shared images, 2DTS still attains the highest Mip-NeRF~360 PSNR among the evaluated methods.

\begin{table}[tb]
\centering
\setlength{\tabcolsep}{3.5pt}
\resizebox{\linewidth}{!}{%
\begin{tabular}{@{}ll|ccc|ccc|ccc@{}}
\toprule
\multirow{2}{*}{\textbf{Dataset}} & \multirow{2}{*}{\textbf{Method}} & \multicolumn{3}{c|}{\textbf{Paper}} & \multicolumn{3}{c|}{\textbf{Public}} & \multicolumn{3}{c}{\textbf{MeshSuite}} \\
& & PSNR $\uparrow$ & SSIM $\uparrow$ & LPIPS $\downarrow$ & PSNR $\uparrow$ & SSIM $\uparrow$ & LPIPS $\downarrow$ & PSNR $\uparrow$ & SSIM $\uparrow$ & LPIPS $\downarrow$ \\
\midrule
\multirow{4}{*}{Mip-NeRF~360}
& 2DTS & \textbf{28.18} & \textbf{0.842} & 0.218 & \textbf{28.51} & \textbf{0.851} & 0.215 & \textbf{28.16}$^\S$ & \textbf{0.841}$^\S$ & \textbf{0.215}$^\S$ \\
& Triangle-Splatting & 27.17 & 0.814 & \textbf{0.192} & 27.20 & 0.813 & \textbf{0.190} & 27.13 & 0.812 & 0.227 \\
& MeshSplatting & 24.78 & 0.728 & 0.310 & 24.78 & 0.728 & 0.310 & 24.72 & 0.729 & 0.365 \\
& DiffSoup & 24.76 & 0.748 & 0.204 & 23.69 & 0.691 & 0.241 & 23.54 & 0.689 & 0.322 \\
\midrule
\multirow{4}{*}{Tanks and Temples}
& 2DTS & \textbf{23.39} & 0.853 & 0.204 & \textbf{23.44} & 0.853 & 0.171 & \textbf{23.42} & 0.853 & 0.203 \\
& Triangle-Splatting & 23.14 & \textbf{0.857} & \textbf{0.143} & 23.09 & \textbf{0.856} & \textbf{0.144} & 23.08 & \textbf{0.856} & \textbf{0.171} \\
& MeshSplatting & 20.52 & 0.745 & 0.287 & 20.58 & 0.758 & 0.272 & 20.53 & 0.756 & 0.319 \\
& DiffSoup & -- & -- & -- & 18.38 & 0.671 & 0.299 & 17.45 & 0.645 & 0.387 \\
\bottomrule
\end{tabular}}
\caption{Reproduction comparisons on Mip-NeRF~360 \citep{mipnerf360} and Tanks and Temples \citep{tanksandtemples}. Bold indicates the best result within each dataset and implementation. DiffSoup does not report Tanks and Temples results in its paper. $^\S$Trained and evaluated with 2DTS's original image pipeline; on the shared images used in all other tables, 2DTS reaches 27.25 / 0.812 / 0.245.}
\label{tab:reproducibility_comparison}
\end{table}

\section{Structural audit details and per-scene results}
\label{app:topology_metrics}
Sec.~\ref{sec:deployment_gap} defines the structural metrics. For connected-component analysis, we exclude faces with repeated vertex indices from the face-adjacency graph and apply no component-size threshold. Tabs.~\ref{tab:mesh_connectivity_details} and~\ref{tab:mesh_connectivity_dtu_details} report per-scene results for all four datasets. Each mean row is the arithmetic average of scene-level metrics; ratios are rounded to three decimal places, valence to two, and mean component counts to the nearest integer. The main-text summary is in Tab.~\ref{tab:mesh_connectivity}.
We highlight the following key findings:

\textbf{Large, unbounded scenes.}
Mip-NeRF~360 and Tanks and Temples exhibit comparable vertex reuse ($|\mathbf{V}|/|\mathbf{F}| \approx 0.52$) and non-manifold ratios (18\% NM edges, 65$\sim$66\% NM vertices). The dataset-level results in Tab.~\ref{tab:mesh_connectivity} show LCC face fractions of 50.3\% and 66.8\%, respectively, but LCC area fractions of only 17.4\% and 17.6\%. The substantial difference between face and area fractions underscores the need to evaluate both measures: most surface area remains outside the LCC, with mean component counts of 286.6k and 171.7k, respectively. Per-scene LCC area fractions range from 3.4\% in \emph{stump} to 46.5\% in \emph{room}; on Tanks and Temples, they are 13.7\% in \emph{truck} and 21.5\% in \emph{train} (Tab.~\ref{tab:mesh_connectivity_details}).

\begin{table}[tb]
\centering
\footnotesize
\textbf{(a) Mip-NeRF~360}\par\smallskip
\begin{tabular*}{\linewidth}{@{\extracolsep{\fill}}lcccccccc@{}}
\toprule
\multirow{2}{*}{\textbf{Scene}} &
\multicolumn{2}{c}{\textbf{Local reuse}} &
\multicolumn{3}{c}{\textbf{Local structure}} &
\multicolumn{3}{c}{\textbf{Global connectivity}} \\
\cmidrule(lr){2-3}\cmidrule(lr){4-6}\cmidrule(l){7-9}
& \textbf{V/F} & \textbf{Val.}
& \shortstack{\textbf{Bnd.}\\\textbf{edges}}
& \shortstack{\textbf{NM}\\\textbf{edges} $\downarrow$}
& \shortstack{\textbf{NM}\\\textbf{verts.} $\downarrow$}
& \shortstack{\textbf{LCC}\\\textbf{faces} $\uparrow$}
& \shortstack{\textbf{LCC}\\\textbf{area} $\uparrow$}
& \textbf{Comp.} \\
\midrule
bicycle & 0.573 & 5.90 & 0.453 & 0.165 & 0.574 & 0.451 & 0.092 & 305,000 \\
flowers & 0.624 & 5.62 & 0.485 & 0.144 & 0.553 & 0.377 & 0.049 & 351,415 \\
garden & 0.476 & 6.85 & 0.427 & 0.196 & 0.700 & 0.605 & 0.286 & 260,582 \\
stump & 0.571 & 5.92 & 0.453 & 0.165 & 0.569 & 0.546 & 0.034 & 339,421 \\
treehill & 0.594 & 5.77 & 0.463 & 0.157 & 0.565 & 0.394 & 0.056 & 343,105 \\
room & 0.465 & 7.22 & 0.461 & 0.184 & 0.752 & 0.670 & 0.465 & 236,839 \\
counter & 0.464 & 7.08 & 0.456 & 0.197 & 0.731 & 0.573 & 0.279 & 238,687 \\
kitchen & 0.408 & 7.56 & 0.404 & 0.238 & 0.746 & 0.701 & 0.264 & 190,968 \\
bonsai & 0.536 & 6.46 & 0.481 & 0.157 & 0.680 & 0.214 & 0.038 & 313,499 \\
\midrule
\textbf{Mean} & \textbf{0.523} & \textbf{6.49} & \textbf{0.454} & \textbf{0.178} & \textbf{0.652} & \textbf{0.503} & \textbf{0.174} & \textbf{286,613} \\
\bottomrule
\end{tabular*}
\par\smallskip
\textbf{(b) Tanks and Temples}\par\smallskip
\begin{tabular*}{\linewidth}{@{\extracolsep{\fill}}lcccccccc@{}}
\toprule
\multirow{2}{*}{\textbf{Scene}} &
\multicolumn{2}{c}{\textbf{Local reuse}} &
\multicolumn{3}{c}{\textbf{Local structure}} &
\multicolumn{3}{c}{\textbf{Global connectivity}} \\
\cmidrule(lr){2-3}\cmidrule(lr){4-6}\cmidrule(l){7-9}
& \textbf{V/F} & \textbf{Val.}
& \shortstack{\textbf{Bnd.}\\\textbf{edges}}
& \shortstack{\textbf{NM}\\\textbf{edges} $\downarrow$}
& \shortstack{\textbf{NM}\\\textbf{verts.} $\downarrow$}
& \shortstack{\textbf{LCC}\\\textbf{faces} $\uparrow$}
& \shortstack{\textbf{LCC}\\\textbf{area} $\uparrow$}
& \textbf{Comp.} \\
\midrule
truck & 0.538 & 6.26 & 0.465 & 0.178 & 0.630 & 0.641 & 0.137 & 162,377 \\
train & 0.495 & 6.72 & 0.458 & 0.187 & 0.700 & 0.695 & 0.215 & 180,929 \\
\midrule
\textbf{Mean} & \textbf{0.516} & \textbf{6.49} & \textbf{0.461} & \textbf{0.182} & \textbf{0.665} & \textbf{0.668} & \textbf{0.176} & \textbf{171,653} \\
\bottomrule
\end{tabular*}
\caption{Per-scene mesh diagnostics for MeshSplatting \citep{meshsplatting} on large, unbounded scenes: (a) Mip-NeRF~360 and (b) Tanks and Temples. Val.: valence; Bnd.: boundary; NM: non-manifold; Comp.: edge-connected components.}
\label{tab:mesh_connectivity_details}
\label{tab:mesh_connectivity_tnt_details}
\end{table}

\clearpage

\begin{table}[t]
\centering
\footnotesize
\setlength{\tabcolsep}{2.5pt}
\renewcommand{\arraystretch}{0.93}
\textbf{(a) DTU}\par\smallskip
\begin{tabular*}{\linewidth}{@{\extracolsep{\fill}}lcccccccc@{}}
\toprule
\multirow{2}{*}{\textbf{Scene}} &
\multicolumn{2}{c}{\textbf{Local reuse}} &
\multicolumn{3}{c}{\textbf{Local structure}} &
\multicolumn{3}{c}{\textbf{Global connectivity}} \\
\cmidrule(lr){2-3}\cmidrule(lr){4-6}\cmidrule(l){7-9}
& \textbf{V/F} & \textbf{Val.}
& \shortstack{\textbf{Bnd.}\\\textbf{edges}}
& \shortstack{\textbf{NM}\\\textbf{edges} $\downarrow$}
& \shortstack{\textbf{NM}\\\textbf{verts.} $\downarrow$}
& \shortstack{\textbf{LCC}\\\textbf{faces} $\uparrow$}
& \shortstack{\textbf{LCC}\\\textbf{area} $\uparrow$}
& \textbf{Comp.} \\
\midrule
scan24 & 0.427 & 7.35 & 0.421 & 0.225 & 0.725 & 0.872 & 0.814 & 23,229 \\
scan37 & 0.365 & 8.13 & 0.418 & 0.278 & 0.799 & 0.929 & 0.891 & 22,489 \\
scan40 & 0.506 & 6.47 & 0.420 & 0.180 & 0.645 & 0.753 & 0.691 & 24,766 \\
scan55 & 0.478 & 6.72 & 0.405 & 0.190 & 0.664 & 0.829 & 0.792 & 16,811 \\
scan63 & 0.383 & 7.94 & 0.395 & 0.248 & 0.786 & 0.647 & 0.595 & 10,646 \\
scan65 & 0.435 & 7.34 & 0.444 & 0.216 & 0.738 & 0.859 & 0.792 & 18,976 \\
scan69 & 0.436 & 7.21 & 0.419 & 0.220 & 0.719 & 0.874 & 0.837 & 18,409 \\
scan83 & 0.394 & 7.81 & 0.396 & 0.233 & 0.772 & 0.919 & 0.882 & 7,366 \\
scan97 & 0.429 & 7.34 & 0.434 & 0.222 & 0.727 & 0.863 & 0.826 & 16,921 \\
scan105 & 0.449 & 7.02 & 0.392 & 0.202 & 0.695 & 0.856 & 0.808 & 13,732 \\
scan106 & 0.425 & 7.50 & 0.428 & 0.207 & 0.735 & 0.837 & 0.773 & 17,091 \\
scan110 & 0.388 & 7.76 & 0.399 & 0.255 & 0.761 & 0.903 & 0.870 & 8,979 \\
scan114 & 0.379 & 7.90 & 0.399 & 0.259 & 0.760 & 0.902 & 0.860 & 13,321 \\
scan118 & 0.486 & 6.61 & 0.397 & 0.186 & 0.658 & 0.826 & 0.791 & 11,679 \\
scan122 & 0.434 & 7.28 & 0.409 & 0.211 & 0.721 & 0.887 & 0.861 & 10,065 \\
\midrule
\textbf{Mean} & \textbf{0.428} & \textbf{7.36} & \textbf{0.412} & \textbf{0.222} & \textbf{0.727} & \textbf{0.850} & \textbf{0.806} & \textbf{15,632} \\
\bottomrule
\end{tabular*}
\par\smallskip
\textbf{(b) NeRF-Synthetic}\par\smallskip
\begin{tabular*}{\linewidth}{@{\extracolsep{\fill}}lcccccccc@{}}
\toprule
\multirow{2}{*}{\textbf{Scene}} &
\multicolumn{2}{c}{\textbf{Local reuse}} &
\multicolumn{3}{c}{\textbf{Local structure}} &
\multicolumn{3}{c}{\textbf{Global connectivity}} \\
\cmidrule(lr){2-3}\cmidrule(lr){4-6}\cmidrule(l){7-9}
& \textbf{V/F} & \textbf{Val.}
& \shortstack{\textbf{Bnd.}\\\textbf{edges}}
& \shortstack{\textbf{NM}\\\textbf{edges} $\downarrow$}
& \shortstack{\textbf{NM}\\\textbf{verts.} $\downarrow$}
& \shortstack{\textbf{LCC}\\\textbf{faces} $\uparrow$}
& \shortstack{\textbf{LCC}\\\textbf{area} $\uparrow$}
& \textbf{Comp.} \\
\midrule
chair & 0.330 & 8.86 & 0.367 & 0.278 & 0.891 & 0.959 & 0.923 & 8,558 \\
drums & 0.335 & 8.82 & 0.387 & 0.271 & 0.886 & 0.469 & 0.523 & 11,856 \\
ficus & 0.398 & 7.32 & 0.367 & 0.282 & 0.798 & 0.318 & 0.621 & 4,743 \\
hotdog & 0.303 & 8.94 & 0.316 & 0.333 & 0.882 & 0.951 & 0.951 & 7,775 \\
lego & 0.312 & 9.21 & 0.367 & 0.293 & 0.916 & 0.973 & 0.951 & 17,679 \\
materials & 0.368 & 8.34 & 0.423 & 0.248 & 0.861 & 0.089 & 0.084 & 19,766 \\
mic & 0.340 & 8.67 & 0.392 & 0.274 & 0.866 & 0.670 & 0.701 & 5,909 \\
ship & 0.323 & 8.85 & 0.356 & 0.295 & 0.883 & 0.960 & 0.954 & 14,265 \\
\midrule
\textbf{Mean} & \textbf{0.339} & \textbf{8.63} & \textbf{0.372} & \textbf{0.284} & \textbf{0.873} & \textbf{0.674} & \textbf{0.713} & \textbf{11,319} \\
\bottomrule
\end{tabular*}
\caption{Per-scene mesh diagnostics for MeshSplatting \citep{meshsplatting} on object-centric datasets: (a) DTU and (b) NeRF-Synthetic. Val.: valence; Bnd.: boundary; NM: non-manifold; Comp.: edge-connected components.}
\label{tab:mesh_connectivity_dtu_details}
\label{tab:mesh_connectivity_nerf_details}
\end{table}

\textbf{DTU results and local defects.} DTU has the highest mean LCC face and area fractions among the four datasets, at 85.0\% and 80.6\%, respectively (Tab.~\ref{tab:mesh_connectivity}). Nevertheless, 19.4\% of surface area remains outside the LCC, and the mean component count is 15.6k. Non-manifold-edge and non-manifold-vertex ratios average 22.2\% and 72.7\%, respectively. Across DTU scenes, LCC area fractions range from 59.5\% in \emph{scan63} to 89.1\% in \emph{scan37}, while component counts range from 7,366 to 24,766 (Tab.~\ref{tab:mesh_connectivity_dtu_details}). Concentrated component mass therefore does not preclude fragmentation or local topological defects in the optimized mesh.

\textbf{NeRF-Synthetic results and high vertex reuse.} On NeRF-Synthetic, LCC face and area fractions average 67.4\% and 71.3\%, respectively, with a mean component count of 11.3k (Tab.~\ref{tab:mesh_connectivity}). Non-manifold-edge ratios average 28.4\% (peaking at 33.3\% in \emph{hotdog}), and NM-vertex ratios average 87.3\% (91.6\% in \emph{lego}). Scene-level connectivity varies substantially: \emph{chair}, \emph{hotdog}, \emph{lego}, and \emph{ship} each place over 95\% of faces and 92\% of area in the LCC, whereas \emph{materials} places only 8.9\% of faces and 8.4\% of area there (Tab.~\ref{tab:mesh_connectivity_nerf_details}). These fractions describe component concentration; they do not by themselves distinguish legitimate scene separation from reconstruction-induced fragmentation. Although vertex reuse is elevated ($|\mathbf{V}|/|\mathbf{F}| = 0.339$, mean valence 8.63), it coexists with high non-manifold ratios and 28.7\% of surface area outside the LCC rather than ensuring a connected, regular 2-manifold surface.

\textbf{Structural audit analysis and practical implications.} Across all four benchmarks, our geometric audit highlights three critical structural challenges:
(1)~\emph{Non-Manifold Structure}: The audited MeshSplatting exports exhibit pervasive non-manifold edges and vertices, with dataset mean ratios of 18$\sim$28\% and 65$\sim$87\%, respectively.
(2)~\emph{Surface Fragmentation}: The LCC encompasses only 17.4$\sim$17.6\% of surface area on the large-scene datasets, compared with 71.3$\sim$80.6\% on the object-centric datasets (Tab.~\ref{tab:mesh_connectivity}). The remaining area lies in other components; mean total component counts range from 11.3k on NeRF-Synthetic to 286.6k on Mip-NeRF~360.
(3)~\emph{Downstream Usability}: These defects motivate additional validation for workflows such as collision handling, UV unwrapping, and simplification. The structural audit does not directly measure performance in these tasks but it reveals that \textit{novel-view fidelity does not ensure structural usability}.

\section{Deployment Results and Resource Measurements}
\label{app:additional_results}

The following sections provide the deployment-gap visualization and scene-level measurements supporting Sec.~\ref{sec:renderer_deployment}, together with NeRF-Synthetic deployment results (Sec.~\ref{app:nerf_synthetic_deployment}). The primary deployment comparison is reported in main-text Tab.~\ref{tab:mip360_renderer_comparison}. Main-text Tabs.~\ref{tab:native_summary} and~\ref{tab:dtu_chamfer_per_scene} report standardized NVS quality and computational cost on three datasets and surface accuracy on DTU, respectively; reproduction comparisons are documented in Appendix Sec.~\ref{app:reproduction}.

\subsection{Deployment Gaps and Cross-Setting Trajectories}
\label{app:deployment_gaps}

\begin{figure}[t]
\centering
\includegraphics[width=\linewidth]{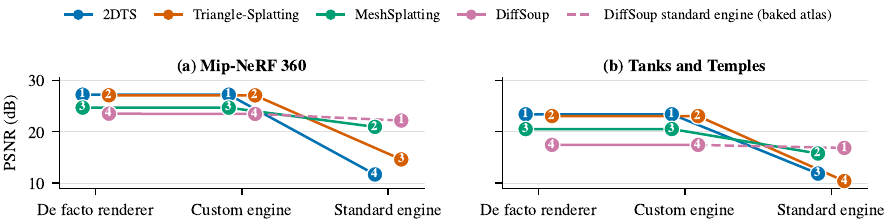}
\caption{Deployment gaps for the four triangle- and mesh-based methods on (a) Mip-NeRF~360 and (b) Tanks and Temples. Each line follows one method across the de facto renderer, custom engine deployment, and standard engine deployment; marker numbers give the rank within each setting. DiffSoup's standard engine deployment uses its baked texture atlas (dashed).}
\label{fig:deployment_gap}
\end{figure}

Fig.~\ref{fig:deployment_gap} visualizes the PSNR trajectories of the four studied triangle- and mesh-based NVS representations across the de facto renderer, standard engine deployment, and custom engine deployment on Mip-NeRF~360 and Tanks and Temples.

\textbf{Cross-setting trajectory analysis.} Across both deployment benchmarks, the visual quality trajectories reveal a consistent pattern:
(1)~\emph{Severe Deployment Degradation for Triangle Soups}: Unconstrained triangle soup methods (2DTS and Triangle-Splatting) rely heavily on soft blending kernels and continuous opacity to represent fine structures and semi-transparent boundaries. When forced into standard engine deployment—where triangles are treated as fully opaque surfaces with hardware Z-buffering—these methods suffer dataset-mean PSNR drops of 11.58$\sim$15.58~dB relative to the de facto renderer. In scenes with complex foliage or fine geometric thinness (\textit{e.g.}, \emph{treehill} or \emph{bonsai}), overlapping transparent triangles collapse into severe opaque, jagged polygon shards (see Fig.~\ref{fig:qualitative_deployment_2dts}).
(2)~\emph{Surface-Constrained Resilience in MeshSplatting}: In contrast, MeshSplatting exhibits far greater resilience under standard engine deployment, dropping by only 3.76~dB on Mip-NeRF~360 (20.96~dB vs.~24.72~dB) and 4.77~dB on Tanks and Temples (15.76~dB vs.~20.53~dB). Because its geometry is constrained to an indexed mesh surface, disabling alpha compositing causes significantly fewer occluding artifacts, allowing it to lead the three vertex-colored methods under standard engine deployment despite lower initial de facto renderer fidelity.
(3)~\emph{Smallest Deployment Gap for DiffSoup}: DiffSoup already uses binary opacity without depth-sorted alpha compositing (Tab.~\ref{tab:design_features}), and its baked texture atlas stays within 1.34~dB of the de facto renderer on Mip-NeRF~360 (22.20~dB vs.~23.54~dB) and 0.60~dB on Tanks and Temples (16.85~dB vs.~17.45~dB), the smallest deployment gaps among the four methods.
(4)~\emph{Recovery under Custom Engine Deployment}: Custom engine deployment returns all four methods to within 0.05~dB of their de facto renderers on both datasets (App. Sec.~\ref{app:custom_renderers}).

\subsection{Mip-NeRF~360 Per-Scene Results}
\label{app:mip360_unity_resources}

Tab.~\ref{tab:mip360_per_scene_quality} reports the per-scene image quality and FPS behind the Mip-NeRF~360 means in Tab.~\ref{tab:mip360_renderer_comparison}(a).

\begin{table}[!t]
\centering
\scriptsize
\setlength{\tabcolsep}{2pt}
\begin{tabular*}{\linewidth}{@{\extracolsep{\fill}}ll*{10}{c}@{}}
\toprule
\textbf{Setting} & \textbf{Method} & \emph{bicycle} & \emph{flowers} & \emph{garden} & \emph{stump} & \emph{treehill} & \emph{room} & \emph{counter} & \emph{kitchen} & \emph{bonsai} & \textbf{Mean} \\
\midrule
\multicolumn{12}{c}{\textit{PSNR $\uparrow$}} \\
\midrule
\multirow{4}{*}{Std.} & 2DTS & 10.08 & 9.46 & 10.91 & 14.19 & 10.87 & 15.72 & 12.53 & 11.45 & 9.78 & 11.67 \\
 & Triangle-Splatting & 12.73 & 10.76 & 13.27 & 14.88 & 12.84 & 18.20 & 16.26 & 16.39 & 16.10 & 14.61 \\
 & MeshSplatting & 20.73 & 17.98 & 20.53 & 22.77 & 19.46 & 23.76 & 21.30 & 20.52 & 21.61 & 20.96 \\
 & DiffSoup & 20.70 & 18.13 & 21.71 & 22.23 & 19.15 & 26.90 & 23.47 & 22.43 & 25.11 & 22.20 \\
\cmidrule(l){2-12}
\multirow{4}{*}{Cust.} & 2DTS & 24.97 & 21.27 & 27.18 & 26.10 & 21.95 & 31.62 & 29.26 & 30.23 & 32.65 & 27.25 \\
 & Triangle-Splatting & 24.85 & 20.68 & 27.15 & 26.16 & 21.69 & 31.14 & 28.91 & 31.32 & 32.04 & 27.11 \\
 & MeshSplatting & 23.02 & 19.21 & 24.70 & 24.87 & 20.47 & 28.28 & 26.34 & 27.31 & 28.20 & 24.71 \\
 & DiffSoup & 21.18 & 18.25 & 23.81 & 22.62 & 19.33 & 28.75 & 25.51 & 25.49 & 26.44 & 23.49 \\
\midrule
\multicolumn{12}{c}{\textit{SSIM $\uparrow$}} \\
\midrule
\multirow{4}{*}{Std.} & 2DTS & 0.205 & 0.173 & 0.239 & 0.258 & 0.236 & 0.486 & 0.387 & 0.333 & 0.370 & 0.299 \\
 & Triangle-Splatting & 0.127 & 0.096 & 0.153 & 0.134 & 0.148 & 0.483 & 0.372 & 0.349 & 0.404 & 0.252 \\
 & MeshSplatting & 0.508 & 0.380 & 0.582 & 0.560 & 0.458 & 0.710 & 0.654 & 0.630 & 0.668 & 0.572 \\
 & DiffSoup & 0.489 & 0.397 & 0.660 & 0.547 & 0.483 & 0.867 & 0.772 & 0.785 & 0.843 & 0.649 \\
\cmidrule(l){2-12}
\multirow{4}{*}{Cust.} & 2DTS & 0.759 & 0.617 & 0.864 & 0.758 & 0.626 & 0.925 & 0.912 & 0.908 & 0.945 & 0.813 \\
 & Triangle-Splatting & 0.764 & 0.609 & 0.863 & 0.757 & 0.605 & 0.926 & 0.910 & 0.929 & 0.946 & 0.812 \\
 & MeshSplatting & 0.638 & 0.491 & 0.760 & 0.685 & 0.538 & 0.872 & 0.845 & 0.858 & 0.877 & 0.729 \\
 & DiffSoup & 0.510 & 0.423 & 0.693 & 0.566 & 0.485 & 0.898 & 0.827 & 0.844 & 0.870 & 0.679 \\
\midrule
\multicolumn{12}{c}{\textit{LPIPS $\downarrow$}} \\
\midrule
\multirow{4}{*}{Std.} & 2DTS & 0.724 & 0.685 & 0.697 & 0.649 & 0.668 & 0.640 & 0.688 & 0.761 & 0.748 & 0.696 \\
 & Triangle-Splatting & 0.673 & 0.684 & 0.639 & 0.670 & 0.682 & 0.629 & 0.659 & 0.672 & 0.648 & 0.662 \\
 & MeshSplatting & 0.470 & 0.508 & 0.370 & 0.450 & 0.537 & 0.505 & 0.503 & 0.475 & 0.534 & 0.484 \\
 & DiffSoup & 0.467 & 0.492 & 0.325 & 0.452 & 0.505 & 0.270 & 0.290 & 0.245 & 0.244 & 0.366 \\
\cmidrule(l){2-12}
\multirow{4}{*}{Cust.} & 2DTS & 0.230 & 0.315 & 0.122 & 0.257 & 0.355 & 0.270 & 0.242 & 0.170 & 0.237 & 0.244 \\
 & Triangle-Splatting & 0.217 & 0.320 & 0.119 & 0.252 & 0.336 & 0.241 & 0.217 & 0.140 & 0.201 & 0.227 \\
 & MeshSplatting & 0.396 & 0.454 & 0.250 & 0.359 & 0.491 & 0.352 & 0.353 & 0.279 & 0.358 & 0.366 \\
 & DiffSoup & 0.430 & 0.464 & 0.272 & 0.415 & 0.477 & 0.223 & 0.238 & 0.194 & 0.212 & 0.325 \\
\midrule
\multicolumn{12}{c}{\textit{FPS $\uparrow$}} \\
\midrule
\multirow{4}{*}{Std.} & 2DTS & 1665 & 1596 & 1511 & 1556 & 1539 & 1565 & 1601 & 1527 & 1438 & 1555 \\
 & Triangle-Splatting & 1297 & 1247 & 1159 & 1355 & 1470 & 1539 & 1379 & 1505 & 1469 & 1380 \\
 & MeshSplatting & 940 & 1016 & 735 & 914 & 948 & 866 & 986 & 907 & 868 & 909 \\
 & DiffSoup & 1779 & 1507 & 1667 & 1687 & 1641 & 1664 & 1706 & 1603 & 1609 & 1651 \\
\cmidrule(l){2-12}
\multirow{4}{*}{Cust.} & 2DTS & 282 & 297 & 269 & 293 & 269 & 192 & 138 & 131 & 199 & 230 \\
 & Triangle-Splatting & 211 & 202 & 229 & 240 & 222 & 462 & 314 & 407 & 334 & 291 \\
 & MeshSplatting & 158 & 167 & 125 & 155 & 158 & 149 & 181 & 172 & 143 & 156 \\
 & DiffSoup & 1433 & 1574 & 1292 & 1577 & 1405 & 1764 & 1964 & 1943 & 1994 & 1661 \\
\bottomrule
\end{tabular*}
\caption{Per-scene image quality and end-to-end FPS on Mip-NeRF~360 under standard engine deployment (Std.) and custom engine deployment (Cust.).}
\label{tab:mip360_per_scene_quality}
\end{table}

\textbf{Per-scene image quality.} The dataset-level ordering holds across scenes (Tab.~\ref{tab:mip360_per_scene_quality}). Under standard engine deployment, DiffSoup or MeshSplatting has the highest PSNR on every scene, while 2DTS and Triangle-Splatting stay below 19~dB throughout (9.46$\sim$15.72~dB and 10.76$\sim$18.20~dB, respectively). Under custom engine deployment, 2DTS has the highest PSNR on seven of the nine scenes and Triangle-Splatting on \emph{stump} and \emph{kitchen}. Custom engine deployment throughput is 131$\sim$297~FPS for 2DTS, 202$\sim$462~FPS for Triangle-Splatting, and 125$\sim$181~FPS for MeshSplatting, whereas DiffSoup stays above 1290~FPS on every scene.

\begin{table}[!t]
\centering
\footnotesize
\resizebox{\linewidth}{!}{%
\begin{tabular}{@{}llcccccccc@{}}
\toprule
 &  & \multicolumn{4}{c}{\emph{train}} & \multicolumn{4}{c}{\emph{truck}} \\
\cmidrule(lr){3-6}\cmidrule(l){7-10}
\textbf{Setting} & \textbf{Method} & PSNR $\uparrow$ & SSIM $\uparrow$ & LPIPS $\downarrow$ & FPS $\uparrow$ & PSNR $\uparrow$ & SSIM $\uparrow$ & LPIPS $\downarrow$ & FPS $\uparrow$ \\
\midrule
\multirow{5}{*}{Std.} & 2DTS & 11.18 & 0.310 & 0.678 & 1631 & 12.50 & 0.363 & 0.632 & 1497 \\
 & Triangle-Splatting & 8.56 & 0.264 & 0.729 & 1537 & 12.20 & 0.334 & 0.639 & 1963 \\
 & MeshSplatting & 14.86 & 0.556 & 0.495 & 1255 & 16.65 & 0.641 & 0.415 & 1509 \\
 & DiffSoup & 15.12 & 0.559 & 0.490 & 1658 & 18.58 & 0.692 & 0.349 & 1895 \\
 & 2DGS & 15.35 & 0.504 & 0.537 & 1495 & 16.86 & 0.545 & 0.493 & 1453 \\
\midrule
\multirow{4}{*}{Cust.} & 2DTS & 21.41 & 0.819 & 0.245 & 503 & 25.45 & 0.886 & 0.157 & 390 \\
 & Triangle-Splatting & 21.22 & 0.822 & 0.215 & 454 & 24.91 & 0.888 & 0.130 & 596 \\
 & MeshSplatting & 18.74 & 0.711 & 0.367 & 247 & 22.34 & 0.805 & 0.272 & 345 \\
 & DiffSoup & 15.52 & 0.569 & 0.476 & 2033 & 19.36 & 0.723 & 0.309 & 2195 \\
\bottomrule
\end{tabular}}
\caption{Per-scene quality and end-to-end FPS on Tanks and Temples under standard engine deployment (Std.) and custom engine deployment (Cust.).}
\label{tab:tnt_renderer_comparison}
\end{table}

\subsection{Tanks and Temples Per-Scene Deployment Results}
\label{app:cross_dataset_results}
Tab.~\ref{tab:tnt_renderer_comparison} reports per-scene image fidelity and throughput for \emph{train} and \emph{truck}.

\textbf{Per-scene fidelity and robustness.} Under custom engine deployment, 2DTS leads in PSNR on both scenes (25.45~dB on \emph{truck} and 21.41~dB on \emph{train}), while Triangle-Splatting has the highest SSIM (0.888 and 0.822) and lowest LPIPS (0.130 and 0.215). MeshSplatting matches its de facto renderer (20.54 vs.\ 20.53~dB). Under standard engine deployment, Triangle-Splatting degrades severely on \emph{train} (dropping 12.66~dB from custom engine deployment to 8.56~dB), while MeshSplatting retains 14.86~dB on \emph{train} and 16.65~dB on \emph{truck}, leading the three vertex-colored methods (15.76~dB mean), and DiffSoup's baked texture atlas reaches 15.12~dB and 18.58~dB (16.85~dB mean). Under custom engine deployment, DiffSoup provides the highest throughput, exceeding 2000~FPS (2033$\sim$2195~FPS), albeit with lower visual fidelity than the other three methods (17.44~dB mean).

\subsection{NeRF-Synthetic Deployment Results}
\label{app:nerf_synthetic_deployment}
Tab.~\ref{tab:nerf_synthetic_deployment} extends the engine deployment comparison to the eight object-centric NeRF-Synthetic scenes under the same protocol as Sec.~\ref{sec:renderer_deployment}.

\begin{table}[tbp]
\centering
\footnotesize
\begin{tabular*}{\linewidth}{@{\extracolsep{\fill}}llcccc@{}}
\toprule
\textbf{Setting}  & \textbf{Method}  & \textbf{PSNR} $\uparrow$ & \textbf{SSIM} $\uparrow$ & \textbf{LPIPS} $\downarrow$ & \textbf{FPS} $\uparrow$ \\
\midrule
\multirow{4}{*}{\shortstack[l]{\textbf{De facto}\\\textbf{renderer}}} & 2DTS & \textbf{33.65} & \textbf{0.970} & \textbf{0.034} & 126 \\
 & Triangle-Splatting & 29.87 & 0.956 & 0.055 & 243 \\
 & MeshSplatting & 29.50 & 0.945 & 0.071 & 86 \\
 & DiffSoup & 28.65 & 0.910 & 0.076 & \textbf{371} \\
\midrule
\multirow{4}{*}{\shortstack[l]{\textbf{Standard engine}\\\textbf{deployment}}} & 2DTS & 20.95 & 0.832 & 0.171 & \textbf{1687} \\
 & Triangle-Splatting & 15.59 & 0.754 & 0.248 & 1616 \\
 & MeshSplatting & 25.69 & 0.903 & 0.119 & 1611 \\
 & DiffSoup & \textbf{27.01} & \textbf{0.924} & \textbf{0.092} & 1686 \\
\midrule
\multirow{4}{*}{\shortstack[l]{\textbf{Custom engine}\\\textbf{deployment}}} & 2DTS & \textbf{33.65} & \textbf{0.970} & \textbf{0.034} & 1203 \\
 & Triangle-Splatting & 29.87 & 0.957 & 0.055 & 1690 \\
 & MeshSplatting & 27.19 & 0.915 & 0.110 & 1793 \\
 & DiffSoup & 29.05 & 0.944 & 0.073 & \textbf{2153} \\
\bottomrule
\end{tabular*}
\caption{Mean NVS quality and end-to-end FPS on NeRF-Synthetic across the three evaluation settings. 
}
\label{tab:nerf_synthetic_deployment}
\end{table}

\textbf{Deployment quality.} The ordering observed on the two real-world datasets carries over. Under standard engine deployment, DiffSoup's baked texture atlas achieves the highest quality (27.01~dB PSNR), followed by MeshSplatting (25.69~dB), 2DTS (20.95~dB), and Triangle-Splatting (15.59~dB); the corresponding deployment gaps are 1.64~dB, 3.81~dB, 12.70~dB, and 14.28~dB. Custom engine deployment reproduces the de facto renderer's mean PSNR for 2DTS (33.65~dB) and Triangle-Splatting (29.87~dB). DiffSoup's custom engine deployment reaches 29.05~dB, 0.40~dB above its de facto renderer, consistent with the engine's 2$\times$ MSAA smoothing object silhouettes against the uniform background, which the single-sample de facto renderer does not.

\textbf{Throughput and resources.} On these small scenes, every engine configuration runs at 1203$\sim$2153~FPS, and the throughput cost of custom engine deployment largely disappears: only 2DTS slows down (1203 vs.~1687~FPS, a 1.4$\times$ reduction), while Triangle-Splatting, MeshSplatting, and DiffSoup are faster than under standard engine deployment. The cost of custom engine deployment therefore depends on scene scale and shader cost.

\section{Engine-Oriented Training Variants}
\label{app:training_variants}
The benchmark in Sec.~\ref{sec:experiments} trains 2DTS and Triangle-Splatting with their standard NVS configurations on Mip-NeRF~360, Tanks and Temples, and NeRF-Synthetic. Both methods additionally provide training configurations that target opaque rendering and may therefore improve compatibility with conventional mesh pipelines. We train these variants and deploy them through the same translation layer and engine renderers as their base methods:

\begin{itemize}
	\item \textbf{2DTS (opaque)}: 2DTS trained with its opaque configuration instead of the standard NVS configuration. Under custom engine deployment, the 2DTS renderer draws its primitives as solid triangles.
	\item \textbf{Triangle-Splatting (game engine)}: Triangle-Splatting trained with the official \texttt{train\_game\_engine.py} configuration, which adds opacity regularization and sigma-based pruning to the standard NVS training.
\end{itemize}

Tab.~\ref{tab:variant_summary} compares each variant with its base method across the evaluation settings. De facto renderer results of the variants are available on Mip-NeRF~360 and Tanks and Temples.

\begin{table}[!t]
\centering
\footnotesize
\resizebox{\linewidth}{!}{%
\begin{tabular}{@{}lllcccc@{}}
\toprule
\textbf{Dataset} & \textbf{Setting} & \textbf{Method} & \textbf{PSNR} $\uparrow$ & \textbf{SSIM} $\uparrow$ & \textbf{LPIPS} $\downarrow$ & \textbf{FPS} $\uparrow$ \\
\midrule
\multirow{12}{*}{Mip-NeRF~360} & \multirow{4}{*}{De facto} & 2DTS & 27.25 & 0.812 & 0.245 & 70 \\
 &  & 2DTS (opaque) & 22.83 & 0.691 & 0.438 & 227 \\
 &  & Triangle-Splatting & 27.13 & 0.812 & 0.227 & 94 \\
 &  & Triangle-Splatting (game engine) & 21.72 & 0.571 & 0.491 & 152 \\
\cmidrule(l){2-7}
 & \multirow{4}{*}{Std.} & 2DTS & 11.67 & 0.299 & 0.696 & 1555 \\
 &  & 2DTS (opaque) & 20.30 & 0.602 & 0.490 & 1656 \\
 &  & Triangle-Splatting & 14.61 & 0.252 & 0.662 & 1380 \\
 &  & Triangle-Splatting (game engine) & 18.31 & 0.381 & 0.609 & 1551 \\
\cmidrule(l){2-7}
 & \multirow{4}{*}{Cust.} & 2DTS & 27.25 & 0.813 & 0.244 & 230 \\
 &  & 2DTS (opaque) & 23.17 & 0.707 & 0.419 & 1854 \\
 &  & Triangle-Splatting & 27.11 & 0.812 & 0.227 & 291 \\
 &  & Triangle-Splatting (game engine) & 21.89 & 0.585 & 0.482 & 508 \\
\midrule
\multirow{12}{*}{Tanks and Temples} & \multirow{4}{*}{De facto} & 2DTS & 23.42 & 0.853 & 0.203 & 101 \\
 &  & 2DTS (opaque) & 19.56 & 0.725 & 0.386 & 294 \\
 &  & Triangle-Splatting & 23.08 & 0.856 & 0.171 & 159 \\
 &  & Triangle-Splatting (game engine) & 17.80 & 0.611 & 0.466 & 148 \\
\cmidrule(l){2-7}
 & \multirow{4}{*}{Std.} & 2DTS & 11.84 & 0.337 & 0.655 & 1564 \\
 &  & 2DTS (opaque) & 17.69 & 0.647 & 0.436 & 1550 \\
 &  & Triangle-Splatting & 10.38 & 0.299 & 0.684 & 1750 \\
 &  & Triangle-Splatting (game engine) & 14.56 & 0.420 & 0.614 & 1603 \\
\cmidrule(l){2-7}
 & \multirow{4}{*}{Cust.} & 2DTS & 23.43 & 0.852 & 0.201 & 447 \\
 &  & 2DTS (opaque) & 19.81 & 0.740 & 0.372 & 1931 \\
 &  & Triangle-Splatting & 23.07 & 0.855 & 0.172 & 525 \\
 &  & Triangle-Splatting (game engine) & 17.93 & 0.626 & 0.455 & 1002 \\
\midrule
\multirow{12}{*}{NeRF-Synthetic} & \multirow{4}{*}{De facto} & 2DTS & 33.65 & 0.970 & 0.034 & 126 \\
 &  & 2DTS (opaque) & 27.79 & 0.932 & 0.103 & 597 \\
 &  & Triangle-Splatting & 29.87 & 0.956 & 0.055 & 243 \\
 &  & Triangle-Splatting (game engine) & 19.29 & 0.817 & 0.214 & 369 \\
\cmidrule(l){2-7}
 & \multirow{4}{*}{Std.} & 2DTS & 20.95 & 0.832 & 0.171 & 1687 \\
 &  & 2DTS (opaque) & 25.50 & 0.910 & 0.115 & 1731 \\
 &  & Triangle-Splatting & 15.59 & 0.754 & 0.248 & 1616 \\
 &  & Triangle-Splatting (game engine) & 16.47 & 0.757 & 0.267 & 1679 \\
\cmidrule(l){2-7}
 & \multirow{4}{*}{Cust.} & 2DTS & 33.65 & 0.970 & 0.034 & 1203 \\
 &  & 2DTS (opaque) & 28.63 & 0.939 & 0.095 & 1994 \\
 &  & Triangle-Splatting & 29.87 & 0.957 & 0.055 & 1690 \\
 &  & Triangle-Splatting (game engine) & 19.45 & 0.827 & 0.208 & 1994 \\
\bottomrule
\end{tabular}}
\caption{Engine-oriented training variants and their base methods under the de facto renderer, standard engine deployment (Std.), and custom engine deployment (Cust.), averaged over the scenes of each dataset.}
\label{tab:variant_summary}
\end{table}

\textbf{Standard engine deployment.} Both variants narrow the gap to the standard opaque mesh pipeline. 2DTS (opaque) improves standard engine deployment PSNR over 2DTS by 8.63~dB on Mip-NeRF~360 (20.30~dB vs.~11.67~dB), 5.85~dB on Tanks and Temples (17.69~dB vs.~11.84~dB), and 4.55~dB on NeRF-Synthetic (25.50~dB vs.~20.95~dB). Triangle-Splatting (game engine) improves over Triangle-Splatting by 3.70~dB, 4.18~dB, and 0.88~dB on the three datasets (18.31~dB, 14.56~dB, and 16.47~dB). With these gains, 2DTS (opaque) approaches MeshSplatting on Mip-NeRF~360 (20.96~dB) and NeRF-Synthetic (25.69~dB), and exceeds both MeshSplatting (15.76~dB) and DiffSoup (16.85~dB) on Tanks and Temples.

\textbf{Custom engine deployment.} The variants trade peak fidelity for runtime cost. Relative to the base methods under custom engine deployment, PSNR drops by 3.62$\sim$5.02~dB for 2DTS (opaque) and by 5.14$\sim$10.42~dB for Triangle-Splatting (game engine). In exchange, 2DTS (opaque) runs at 1854~FPS on Mip-NeRF~360 and 1931~FPS on Tanks and Temples, 8.1$\times$ and 4.3$\times$ faster than 2DTS. Triangle-Splatting (game engine) reaches 508~FPS and 1002~FPS on the two datasets, 1.7$\times$ and 1.9$\times$ faster than Triangle-Splatting.

\textbf{Gaps to the de facto renderer.} On Mip-NeRF~360 and Tanks and Temples, custom engine deployment of the variants is 0.13$\sim$0.34~dB above their de facto renderers: 23.17~dB vs.~22.83~dB and 19.81~dB vs.~19.56~dB for 2DTS (opaque), and 21.89~dB vs.~21.72~dB and 17.93~dB vs.~17.80~dB for Triangle-Splatting (game engine). Standard engine deployment trails the de facto renderer by 2.53~dB and 1.87~dB for 2DTS (opaque) and by 3.41~dB and 3.24~dB for Triangle-Splatting (game engine), compared with 11.58$\sim$15.58~dB for the base methods. The removal of view-dependent SH appearance may contribute to these remaining gaps, although the comparison does not isolate individual causes. The de facto renderer PSNR of the variants is itself 3.86$\sim$5.41~dB below that of the base methods, so the smaller gaps do not imply higher absolute rendering quality.

\textbf{Portability gap.} The portability gap between the two engine settings shrinks to 2.12$\sim$3.13~dB for 2DTS (opaque) and 2.98$\sim$3.58~dB for Triangle-Splatting (game engine), compared with 11.59$\sim$15.58~dB and 12.50$\sim$14.28~dB for the base methods. Training toward opaque triangles thus improves compatibility with the standard opaque mesh pipeline and lowers runtime cost, but in these configurations it does not match the image quality that the base methods reach under custom engine deployment.

\section{Qualitative Deployment Analysis of 2DGS Baseline}
\label{app:2dgs_qualitative}

\begin{figure}[t]
\centering
\includegraphics[width=\linewidth]{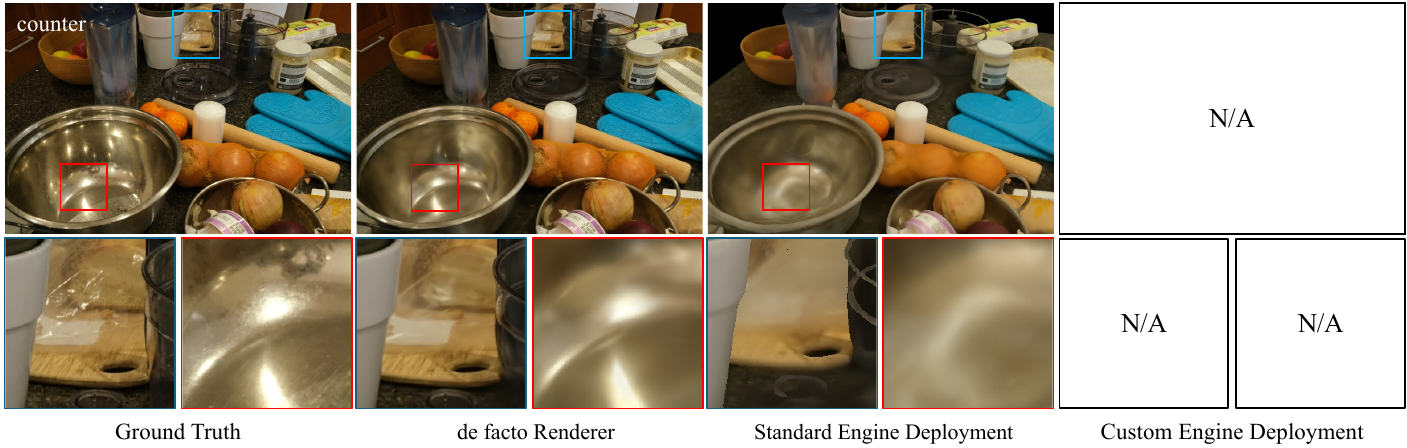}
\caption{Qualitative NVS comparison of 2DGS \citep{2dgs} on the \emph{counter} scene of Mip-NeRF~360. Columns show ground truth, de facto renderer, standard engine deployment, and custom engine deployment. Standard engine deployment uses a TSDF-extracted mesh; custom engine deployment is unsupported under our deployment protocol (\textsc{N/A}).}
\label{fig:qualitative_deployment_2dgs}
\end{figure}

Fig.~\ref{fig:qualitative_deployment_2dgs} supplements the triangle- and mesh-based comparisons in Fig.~\ref{fig:qualitative_deployment_2dts} with the 2DGS \citep{2dgs} reference baseline.

\textbf{Asymmetry with neural feature decoding.} 2DGS and DiffSoup illustrate different compatibility constraints under the evaluated deployment protocol. DiffSoup's neural color decoder has no counterpart in standard engine deployment, where it is replaced by a baked texture atlas (22.20~dB PSNR on Mip-NeRF~360 and 16.85~dB on Tanks and Temples), while custom engine deployment retains it and achieves 23.49~dB and 17.44~dB, respectively (Tab.~\ref{tab:mip360_renderer_comparison}). For 2DGS, TSDF fusion of rendered depth maps produces a triangle mesh with static vertex colors. The exported mesh supports standard engine deployment but does not retain the Gaussian splatting parameters, opacity, and view-dependent appearance needed to reproduce the de facto renderer. Custom engine deployment is therefore marked \textsc{N/A} for this exported asset. This designation concerns the evaluated exported asset; it does not establish that a dedicated Gaussian renderer could not be implemented in a graphics engine.

\textbf{Visual degradation under standard engine deployment.} The static-color opaque mesh extracted from 2DGS does not preserve the appearance of the de facto renderer:
(1)~\emph{Reflective appearance}: In the red inset around the metallic bowl, the de facto renderer retains a localized bright reflection, although it is already softer than the ground truth. Under standard engine deployment, this reflection becomes more diffuse and loses contrast. Static vertex colors cannot reproduce changes in appearance with viewing direction.
(2)~\emph{Fine detail and object boundaries}: The cyan inset around the wrapped cutting board shows blurred texture and less distinct boundaries after mesh conversion. These differences are consistent with losses introduced by surface reconstruction and static-color rendering; this comparison does not isolate the contributions of TSDF discretization, color baking, and opaque visibility.
(3)~\emph{Quantitative contrast}: On Mip-NeRF~360, dataset-mean PSNR falls from 26.79~dB to 12.70~dB, a 14.09~dB decrease, while LPIPS increases from 0.252 to 0.594 (Tab.~\ref{tab:mip360_renderer_comparison}). These dataset-level metrics complement the single-scene visualization and describe the combined effect of mesh conversion and standard engine deployment.

\end{document}